\documentclass[%
 reprint,
 amsmath,amssymb,
 aps,
pre,
english]{revtex4-2}

\usepackage{graphicx}
\usepackage{dcolumn}
\usepackage{bm}
\usepackage{upgreek}

\usepackage{cleveref}   
\usepackage{color}
\usepackage{subcaption}
\usepackage{caption}
\usepackage{soul}
\begin{document}

\title{
Shear flow of frictional spheroids: Comparison between elongated and flattened particles
}
\thanks{© 2025 American Physical Society. This is the author’s accepted manuscript, made available under APS policy for Green Open Access. The published version is available at \url{https://doi.org/10.1103/tj41-6qqk}.}

\author{Jacopo Bilotto}
    \altaffiliation{Institute of Civil Engineering, Institute of Materials,
            \'{E}cole Polytechnique F\'{e}d\'{e}rale de Lausanne (EPFL), CH 1015 Lausanne, 
            Switzerland}
\author{Martin Trulsson}
            \altaffiliation{Computational Chemistry, Lund University, Lund SE-221 00, Sweden}
\author{Jean-François Molinari}
            \altaffiliation{Institute of Civil Engineering, Institute of Materials,
            \'{E}cole Polytechnique F\'{e}d\'{e}rale de Lausanne (EPFL), CH 1015 Lausanne, 
            Switzerland}

\date{October 30 2025}

\begin{abstract}

The rheology of dense granular shear flows is influenced by friction and particle shape.
We investigate numerically the impact of non-spherical particle geometries under shear on packing fraction, stress ratios, velocity fluctuations, force distribution, and dissipation mechanisms, for a wide range of inertial numbers, friction coefficients and aspect ratios.
We obtain a regime diagram for the dissipation which shows that lentil-like (oblate) particles exhibit an extended sliding regime compared to rice-like (prolate) particles with the same degree of eccentricity.
Additionally, we identify non-monotonic behaviour of slightly aspherical particles at low friction, linking it to their higher fluctuating rotational kinetic energy.
We find that angular velocity fluctuations are generally reduced when particles align with the flow, except in highly frictional rolling regimes, where fluctuations collapse onto a power-law distribution and motion becomes less correlated.
Moreover, for realistic friction coefficients power dissipation tends to concentrate along the major axis aligned with the flow, where slip events are more frequent.
We also show that flat particles develop stronger fabric anisotropy than elongated ones, influencing macroscopic stress transmission.
These findings provide new insights into the role of particle shape in granular mechanics, with implications for both industrial and geophysical applications. 

\end{abstract}

\maketitle

\section{Introduction}

Granular materials exhibit complex behavior, making their study fundamental in physics, engineering, and materials science.
Two key factors influencing granular mechanics are the microscopic friction, via surface roughness of the grains \cite{andreotti2013granular}, and particle shape \cite{borzsonyi2013granular}.
For example, elongated and flattened particles can pack more densely than spheres, affecting their jamming transitions and critical bulk properties \cite{delaney2010packing, brito2018universality}, while biaxial compression experiments provide evidence that elongated particles exhibit enhanced interlocking, which further strengthens the granular assembly \cite{azema2010stress}.

The influence of particle shape also plays a crucial role under shear.
Plate-like particles in quasi-static critical state flow show increased shear strength and ordering effects \cite{boton2013quasistatic}.
It was shown that even small degrees of eccentricity lead to considerable alignment under shear, close to the jamming limit \cite{marschall2019orientational}. \\
Similar effects are also present when particle inertia cannot be neglected.
Experimental studies provide direct evidence of alignment of sheared elongated \cite{borzsonyi2012shear, wegner2012alignment, pol2022kinematics} and flattened \cite{fan2024effect} particles in a variety of configurations.
Numerical methods, particularly the discrete element method (DEM) \cite{cundall1979discrete} have confirmed the experimental evidence.\\ 
In simple shear simulations at constant volume fraction, different particle geometries lead to distinct flow behaviors: spheres display well-characterized rheology \cite{campbell2002granular}, while ellipsoids \cite{campbell2011elastic}, rods \cite{guo2012numerical}, and flat disks \cite{guo2013granular} exhibit unique flow properties such as collective entanglement and enhanced stress anisotropy.
Friction between particles reduces alignment, but increase shear-thickening \cite{nath2019rheology}.
Moreover, studies on ellipsoids reveal how alignment affects diffusion under shear \cite{cai2021diffusion}.\\
Under constant applied pressure, the shear behavior changes slightly.
The foundational work on spheres \cite{da2005rheophysics} has set the stage for granular rheology, which was extended to frictionless sphero-cylinders \cite{nagy2017rheology}. 
Investigations of 2D ellipses show that friction significantly affects shear jamming and rheology in a nonmonotonic way \cite{trulsson2018rheology}, while extending these studies to 3D frictional spherocylinders allowed to directly relate normal stress differences to particle-level contacts \cite{nagy2020flow}. 
The anisotropy of contact also leads to an increase of the frictional forces contribution to shear strength for elongated particles \cite{zou2023particle}.

To capture the physics of these shape-dependent behaviors, various modeling approaches have been developed.
The classical \(\mu(I)\) rheology, originally formulated for spheres \cite{jop2006constitutive}, has been extended to frictionless spheroids \cite{nadler2021anisotropic}, to account for kinematic orientation \cite{nadler2018kinematic}. 
The effect of shape has also been incorporated in the kinetic theory of  both frictionless \cite{berziStressesOrientationalOrder2016, berziCollisionalDissipationRate2017} and frictional grains \cite{berziDenseShearingFlows2022, vescoviSimpleGeneralizationKinetic2024}, of non-spherical particles via an effective coefficient of restitution to account for frictional dissipation and both order tensors and elastic stresses. 
These models provide a high degree of accuracy with only a couple of fitting parameters, yet their application relies on the measurements of critical volume fraction, which can be challenging to determine under constant stress conditions \cite{campbellStresscontrolledElasticGranular2005}.

Regime diagrams describing dissipation mechanisms as a function of inertial number and friction have evolved from early studies on 2D granular disks \cite{degiuli2016phase} to more complex dense suspensions \cite{trulsson2017effect} and 3D spherocylinders \cite{nagy2020flow}, which showed an extended sliding regime. 
Yet, the regime diagram for more complex shapes is missing, and predicting it does not appear trivial.

Spatial correlations close to the jamming transition have been analyzed and characterized for both dry granular flows and suspensions
\cite{pouliquen2004velocity, silbert2005vibrations, heussinger2013shear, during2014length, ness2016shear, hexner2018two, singh2020shear, zou2022microscopic}, emphasizing the role they play in correlated clusters.
Despite evidence of spatial correlations playing a key role near jamming, their dependence on particle shape, especially in terms of aspect ratio and out-of-plane dimensions, has not been systematically investigated.

While significant progress has been made in understanding the role of elongation in granular flow, the comparative influence of flatness versus elongation remains underexplored—particularly in frictional, dense regimes.

In this work, we systematically examine the quasi-2D shear flow of frictional spheroids (ellipsoids of symmetry) in the dry dense regime,
focusing on how different aspect ratios influence particle motion and stress transmission. 
By comparing lentil-like (oblate) and rice-like (prolate) grains, we aim to uncover how shape variations dictate sliding behavior, macroscopic flow properties and microscopic interactions.
We focus on comparing the effect of the out of shear plane dimension of particles, in what is predominantly a 2D flow.
Through numerical simulations, we characterize the orientation of the particles and obtain a regime diagram for each particle shape. 
We then analyze the components of the fluctuating kinetic energy, the velocity fluctuations and their spatial correlations, and several other rheological and particle-level quantities, to understand the emergence of different shear-flow regimes. 
The influence of particle shape on the non-linear contribution of macroscopic friction in shear flow deepens the understanding of shape effects in granular mechanics, showing how even in simple quasi-2D flows, the full 3D geometry of the particles plays a role.
This finding could be used to inform and develop new models for non-spherical grains rheology.

\begin{figure*}
    \centering
    \includegraphics[width=0.8\textwidth]{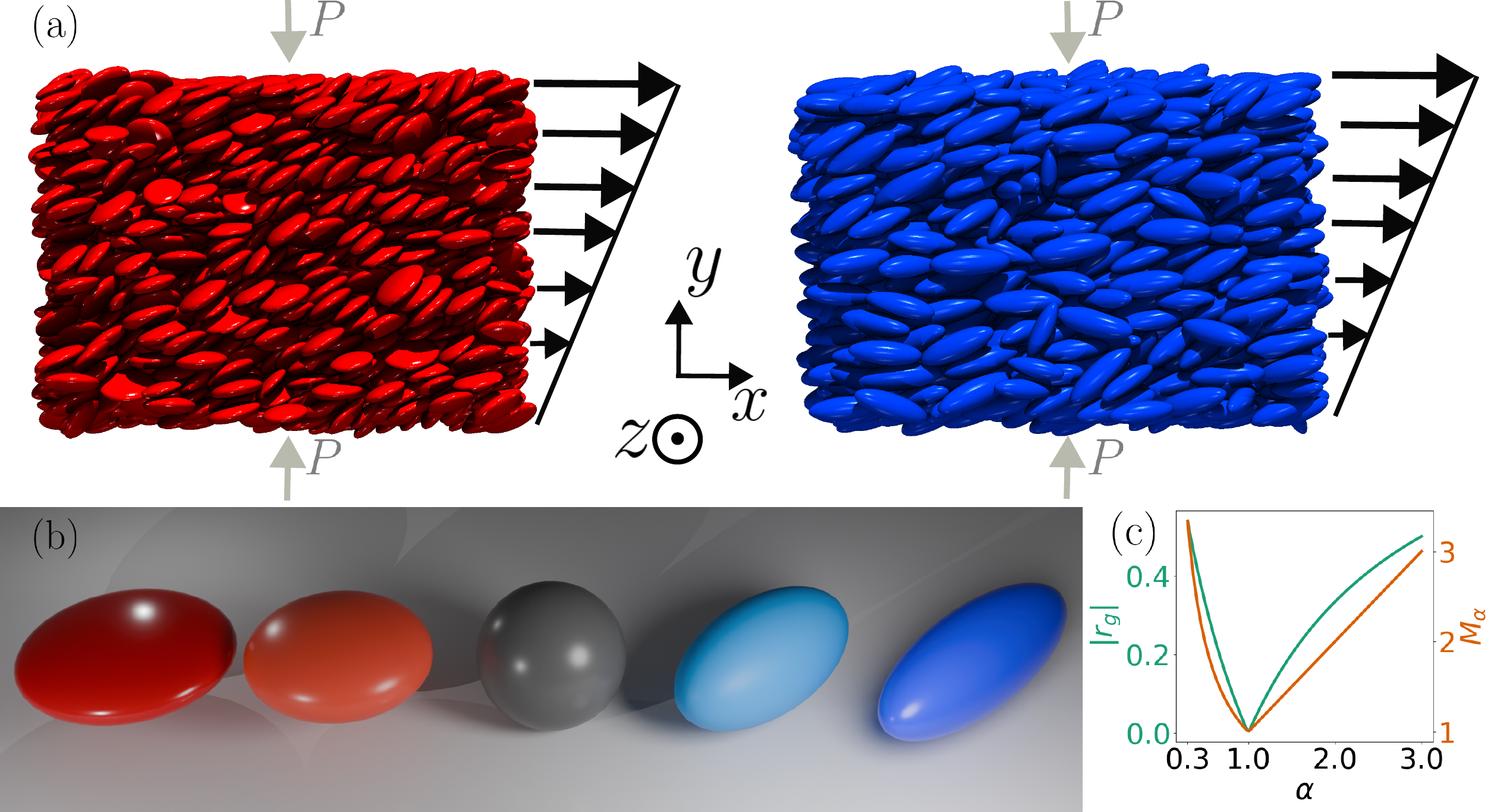}
    \caption{(a) Snapshot of the simulation box.
     Oblate grains $\alpha = 0.33, \, r_g=-0.5$ and prolate ones $\alpha = 3.0, \, r_g=0.5 $ are displayed on the left and right respectively. A constant shear rate in the $xy$ plane, $\dot{\gamma}$, is applied by tilting the box, while the normal stress $\sigma_{yy}=P$ is controlled by shrinking the box in the $y$ direction.
     (b) Illustration of ellipsoids' shapes in this study from flattened to elongated left to right.
    (c) Absolute value of the shape ratio $|r_g|$ and $M_\alpha = \max(\alpha, \alpha^{-1})$ as a function of aspect ratio $\alpha$. 
    }
    \label{fig:sketch}
\end{figure*}

\section{Numerical Protocol}
\label{sec:simulations}
All simulations are performed with the open source software LIGGGHTS \cite{goniva2010open}.
To achieve simple shear flow we use a 3D triclinic box, with periodic boundary conditions in the $x$ and $z$ directions, respectively the flow and vorticity one, and Lees-Edwards \cite{lees1972computer} boundary conditions in the $y$ gradient direction, achieved by continuously tilting the box via the \texttt{fix-deform} command, see Figure \ref{fig:sketch}a and b for a depiction. 
We model spheroids using the superquadric particle model \cite{podlozhnyukEfficientImplementationSuperquadric2016}.
The equation of a spheroid oriented along its principal axes and with its center coinciding with the one of the coordinate system is
\begin{align}
    {\mathcal{X}^2 +\mathcal{Y}^2 \over a^2} + {\mathcal{Z}^2 \over c^2} = 1    \, .
\end{align}
We define the aspect ratio $\alpha \equiv c/a$ and the shape ratio $r_g \equiv (c-a)/(c+a)$ so that prolate and oblate spheroids correspond to $\alpha>1, r_g>0$ and $\alpha<1, r_g<0$ respectively.
We also define $M_\alpha = \max(\alpha, 1/\alpha)$ as another measure of ``sphericity'', which we will use in the text. 
The relationship between $M_\alpha$ and $|r_g|$ is additionally shown in Figure \ref{fig:sketch}c.  

Grains interact with each other through contacts.
Along the contact normal $\mathbf{n}$, we employ 
the Hertzian model \cite{hertz1882beruhrung} with a normal damping, whose value is related to our chosen restitution coefficient \cite{luding2008introduction}.
The tangential interaction follows Mindlin theory \cite{mindlin1953elastic} with a Coulomb friction limit on the tangential force.
The local curvature coefficient for the ellipsoids is chosen as the gaussian curvature coefficient since it yields the best results in terms of normal ellipsoid-ellipsoid interaction \cite{podlozhnyukEfficientImplementationSuperquadric2016}.
The time step is selected as the minimum between $15\%$ of the Hertzian binary collision \cite{johnson1987contact} and $15\%$ of the Rayleigh time.
See Appendix \ref{app:contact} for more details on the contact interaction, and Appendix \ref{app:dissipation} for the negligible effect on varying the time step.

The components of the stress tensor are calculated via the virial stress formula: 
\begin{align}
    \sigma_{ij} =& {1 \over 2\Omega} \sum_{k, l \in V}^N  (r_{j}^{(k)}- r_{j}^{(l)})F_{i}^{(lk)}
\end{align}
where $\Omega$ is the volume of the system, $N$ is the number of particles,
$\mathbf{r}^{(k)}$ the particle position vector and 
$\mathbf{F}^{(lk)}$ the force acting on particle $l$ due to the interaction with particle $k$. 
We neglect the contribution of the velocity fluctuations to the stress, since in the dense flows we are investigating (packing fraction $\phi > 0.5$) it is at least two orders of magnitude smaller than the contact stress \cite{campbell1989stress}, which we also confirmed by computing it in post-processing. 

We control the normal pressure component $\sigma_{yy} = -P$, by assigning a velocity $d L_y/dt = k_p (P_t-\overline{P})$ to the box edges with normals $\pm \mathbf{e}_y$, where ${\overline{P}}$ is the average pressure calculated between the control inputs, $P_t$ is the target vertical pressure, and $k_p$ is the proportional control gain.
We set the target pressure so that the contact stiffness number $\kappa = \min(\alpha, 1) (E/P)^{2/3}
>1.1\cdot 10^3$, where $E$ is the Young modulus of the grains. 
Compared to \cite{da2005rheophysics} we scale the contact stiffness number by the aspect ratio, in order to guarantee a small overlap even for oblate particles.
Doing so ensures we are in the hard grain limit \cite{da2005rheophysics, peyneau2008frictionless} and
the relevant dimensionless number for several rheological properties \cite{favierdecoulombRheologyGranularFlows2017} is the inertial number $I = \dot{\gamma} d_{eq} \sqrt{\rho_p/P}$, where $\dot{\gamma}$ is the shear rate, $\rho_p$ is the particle density, $d_{eq} = 2 a \alpha^{1/3}$ is the equivalent volume sphere diameter.
Nevertheless, finite particle elasticity affects both coordination and volume fraction \cite{chialvoBridgingRheologyGranular2012, favierdecoulombRheologyGranularFlows2017, berziDenseShearingFlows2022, berziGranularFlowsKinetic2024}.
We set the Poisson's ratio $\nu = 0.3$, which sets the ratio of tangential to normal stiffness $k_t/k_n \approx 0.41$, see Appendix \ref{app:contact} for more detail. 

For each simulation, we apply the control input at constant values of strain (rather than strain rate) during the simulation.
We empirically found that, if for every shear step $\dot{\gamma} t =1$ we apply the controller $1000 I^{-1/3}$ times, we keep on average the desired target pressure  for all the simulations reported here.
We also limit the control input so that the box does not shrink more than $10^{-2} L_y$ between consecutive inputs, even though in most cases the value is much smaller.
While applying the control input we also subtract the center of mass average velocity from all particles to avoid drifting of the simulation box. 

We reach a final strain $\dot{\gamma} t =16$, and the averages are computed over more than $1000$ snapshots at constant strain intervals starting at $\dot{\gamma} t =6$ for $I\geq0.01$ and $\dot{\gamma} t =4$ for the other cases, to ensure a steady state is reached for all simulations.
We discard those simulations where the average inertial number measured is more than $10\%$ off from the target one.

$N=2000$ particles are inserted in the box, with approximately $12$ particles per dimension and, with a $ 20\%$ degree of polydispersity for the value of the smallest semi-axis around the mean and none in the aspect ratio.
All simulations start at an initial concentration $\phi_0 = 0.4$ with isotropic random distribution and orientation of grains, equal edges of the box $L_x=L_y=L_z$, and then the box is compressed in the $y$ direction to reach the desired pressure. 
The chosen number of particles is sufficient to characterize the rheology, see Appendix \ref{app:finite_size}.
We set the coefficient of restitution $e=0.1$, insofar at the high packing fractions ($\phi>0.45$) we investigate, its influence in the range $[0.1, 0.9]$ on several rheological quantities is minor \cite{da2005rheophysics, peyneau2008frictionless, guo2013granular, hurleyFrictionInertialGranular2015, degiuli2016phase, favierdecoulombRheologyGranularFlows2017, berziGranularFlowsKinetic2024}. 

We explore different particle shapes $\alpha\in [0.33, 3.0]$ and inertial numbers $I\in [10^{-3}, 10^{-1}]$.
We also vary the microscopic friction coefficient between the particles, $\mu_p \in [0, 10]$.

All the quantities reported are averaged over all particles and all time steps at steady-state, unless stated otherwise.

\section{Results}
\label{sec:results}
Since we simulate a quasi-2D flow with 3D particles it is interesting to compare the results at the same value of $|r_g|$, or equivalently the same of $M_\alpha$, for oblate and prolate particles.
Two spheroids with shape ratios $r_{g, 1} = - r_{g, 2}$, or equivalently $\alpha_1=1/\alpha_2$ would correspond to the same 2D ellipse in the $xy$ shear plane, provided the axis of symmetry lies in that plane, so that a direct comparison between the two is warranted.
We refer to this shapes as ``reciprocal'' in the remainder of the text. \\
In the remainder of this manuscript we will use the same color code as in Figure \ref{fig:sketch}(a) and (c) to distinguish between particle shapes.

We begin by characterizing the orientational ordering and alignment of the particles.
Then, we measure the power dissipation and analyze the regime diagram for non-spherical particles. 
We proceed to explain the non-monotonicity of the regime transition for small eccentricity and we try to explain why the oblate spheroids show and extended sliding regime compared to prolate ones.
To do so, we investigate additional rheological and jamming characteristics, revealing that oblate particles exhibit stronger velocity correlations and fabric anisotropy compared to their prolate counterparts.
Finally we look at the microscopic interactions in the sliding regime where we observe power dissipation clustering on the major axes mostly aligned with the streamlines.

\subsection{Alignment and Rotation}
\label{sec:alignment}
\begin{figure}
    \centering
   \includegraphics[width=\linewidth]{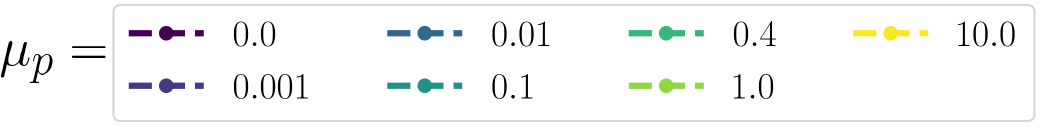}
    \includegraphics[width=\linewidth]{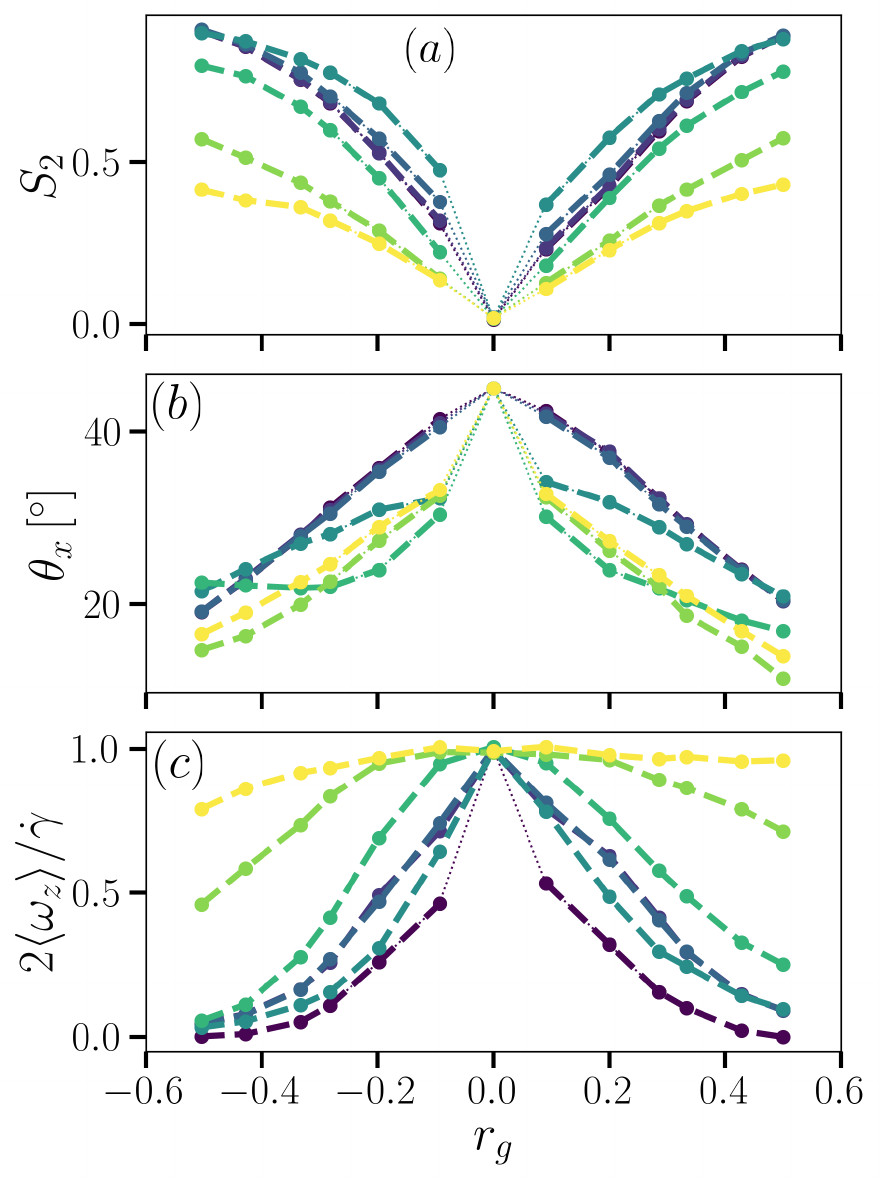}
    \caption{
    (a) Nematic order parameter, (b)  mean angle with the shear flow, and (c) average grain angular velocity in the vorticity direction.
    All data in this figure is measured at $I\approx 0.01$ and is nearly constant over the range of $I$ in this study.
    Dotted lines indicate the behavior towards ill-defined quantities for spheres.}
    \label{fig:S2_theta_omega_z}
\end{figure}

As the first step to characterize the behavior of anisotropic particles, we analyze their orientation. 
The inertia of the particles has a small influence on the orientation of the grains, we therefore report data at a fixed value of $I\approx10^{-2}$ in Figure \ref{fig:S2_theta_omega_z}.\\
Let us define $\mathbf{p}$ the unit vector describing the orientation of the symmetry axis of the spheroid, then the orientational tensor can be defined as 
\begin{align}
    \mathbf{Q} = \int_{\mathbb{S}^2} \left(\mathbf{p} \otimes \mathbf{p} - {\mathbf{I} \over 3}\right) \psi(\mathbf{p}) d \mathbf{p} \, ,
\end{align}
where $\mathbf{I}$ is the identity matrix, and $\psi(\mathbf{p})$ is the probability distribution of $\mathbf{p}$ on the unit sphere, $\mathbb{S}^2$.
The uniaxial nematic order parameter is defined as $S_2 = 3 \lambda_{\text{max}}/2$, where $\lambda_{\text{max}}$ is the maximum eigenvalue of $\mathbf{Q}$.
As shown in Figure \ref{fig:S2_theta_omega_z}a, $S_2$ increases from $0$ for an isotropic distribution when $r_g \to 0$, to values close to $1$ for elongated and flattened particles, where a value $S_2=1$ means that all particles are aligned in the same direction. 
Additionally, we define as weakly anisotropic configurations where $0<S_2<0.35$.

All non-spherical particles exhibit some degree of nematic alignment.
Furthermore, we verified that no persistent smectic phase exists by computing the transverse correlation function of the center of mass density field \cite{marschall2019orientational}, whose rapid decay confirms the absence of long-range translational order.\\
One might expect that microscopic friction broadens the distribution of particle orientations, thereby reducing nematic order.
However, for particles with low to moderate eccentricity, the highest degree of nematic ordering is not observed in the frictionless limit, but at an intermediate friction coefficient of $\mu_p = 0.1$.
This underscores the dual role of friction: while it typically disrupts alignment in the nematic phase, it can promote alignment in the isotropic phase via torque-induced reorientation. 
Finally, notice how for small eccentricity, $r_g<0.1$ the order parameter is within the weakly anisotropic case.
Interestingly, oblate particles align slightly better up to moderate values of friction coefficient $\mu_p \leq 0.4$, but worse at high friction values with respect to their reciprocal prolate ones.\\
As reported in several studies \cite{campbell2011elastic, borzsonyi2012shear, wegner2012alignment, berziStressesOrientationalOrder2016, berziCollisionalDissipationRate2017, nagy2017rheology, trulsson2018rheology, nagy2020flow, cai2021diffusion, pol2022kinematics}
grains do not simply align in the direction of least resistance to the flow, along the streamlines, but at a certain angle in the extension direction.
Defining the mean angle of inclination of the major axis with respect to the flow direction projected in the $xy$ plane, $\theta_x$, \footnote{for oblate ellipsoids since there are are two major axes we track the symmetry axis and then compute the corresponding angle for the major axis as $\pi/2-\theta_x$}, we report data in Figure \ref{fig:S2_theta_omega_z}b.
The higher $|r_g|$ the more the particles orient their major axis parallel to the flow direction.
The effect of friction is also highly non-linear: $\theta_x$ is lower for $0.1<\mu_p<10$ at small degrees of eccentricity, but for high flattening and elongation, the minimum $\theta_x$ is found for high values of friction, when alignment is less intense.
The existence of this alignment angle, known as the Leslie angle \cite{leslieCONSTITUTIVEEQUATIONSANISOTROPIC1966} in liquid crystal theory, is well-established. 
However, its microscopic origins in granular matter are not fully understood, and predicting its value from fundamental particle properties is an open problem.
Recent work \cite{talbotExploringNoisyJeffery2024} has started to illuminate these origins, linking the angle's stability to collisional noise in the shear flow.\\
Another way to look at the emergence of alignment is to track the the average vorticity-component of angular velocity $\langle \omega_z \rangle$,  see Figure \ref{fig:S2_theta_omega_z}c.
When friction is present, spherical particles rotate on average as a fluid particle in shear flow \cite{da2005rheophysics} $\omega_z \approx \dot{\gamma}/2$.
For non-spherical particles, elongation provides a geometric constraint, effectively hampering their rotation.
For all values of friction, oblate particle rotate slightly less than prolate ones for the same degree of eccentricity.
One possible explanation is that while some prolate particles orient their major axis along the vorticity direction, resulting in a log-rolling behavior, this is not observed for oblate particles (data not shown).
This behavior is analogous to that seen in simple shear flow in a dilute viscous fluid, where rotation about the vorticity axis is stable for prolate spheroids when their minor axis aligns with it, but for oblate spheroids, stability occurs when their major axis is aligned instead \cite{hinch1979rotation}.
In the high friction limit, yellow curve, the rotation approaches the one of the corresponding fluid particle, regardless of the shape.
This is due to the lower packing fraction attained in this scenario, see Figure \ref{fig:mu_I}.
To conclude, for non-spherical particles we notice that the average rotation is anticorrelated with the orientational order \cite{marschall2019orientational}, but not directly related to the particles' preferred orientation.

\subsection{Dissipation Regime diagram and fluctuations}

\begin{figure}
    \centering
    \includegraphics[width=\linewidth]{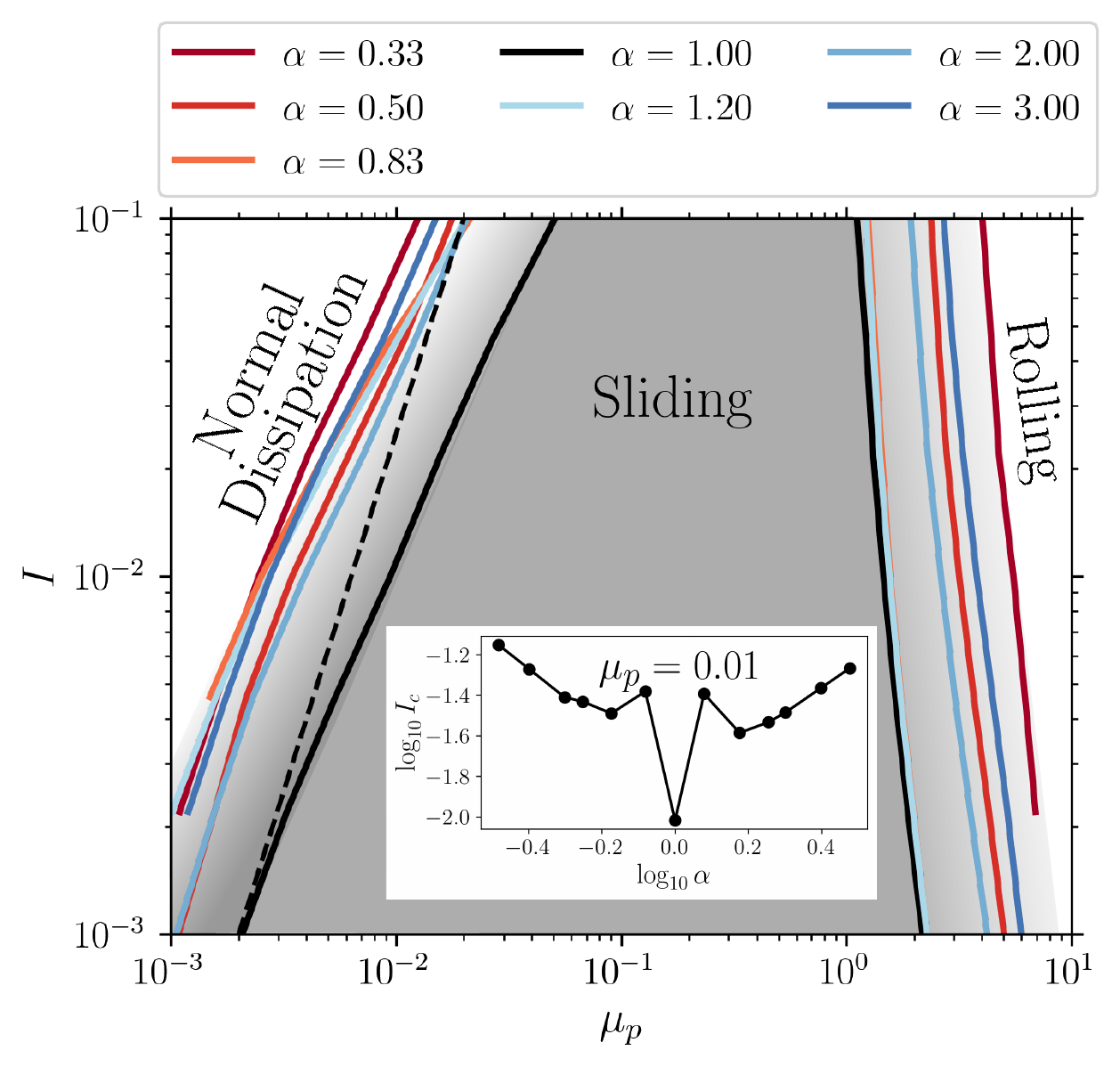}
    \caption{Dissipation regime diagram of granular shear flow as a level set curve where the frictional and normal dissipation are equal.
    Left lines mark the transition between the normal dissipation and the sliding frictional regime, while right ones the transition to the rolling regime, where virtually all contacts are sticking.
    The black dashed line has slope 2 as predicted 
    by \cite{degiuli2016phase} for spheres.
    For all non-spherical particles the sliding regime is wider than for spherical ones, indicating non-spherical particles slide over a larger range of parameters.
    (Inset) Slice of $I_c$ at the normal dissipation-sliding transition for $\mu_p=0.01$, showing non monotonicity in $\alpha$. 
    Slightly aspherical particles display a reduced normal dissipation regime. 
    }
    \label{fig:regime_diagram}
\end{figure}

\begin{figure}
    \centering  
    \includegraphics[width=\linewidth]{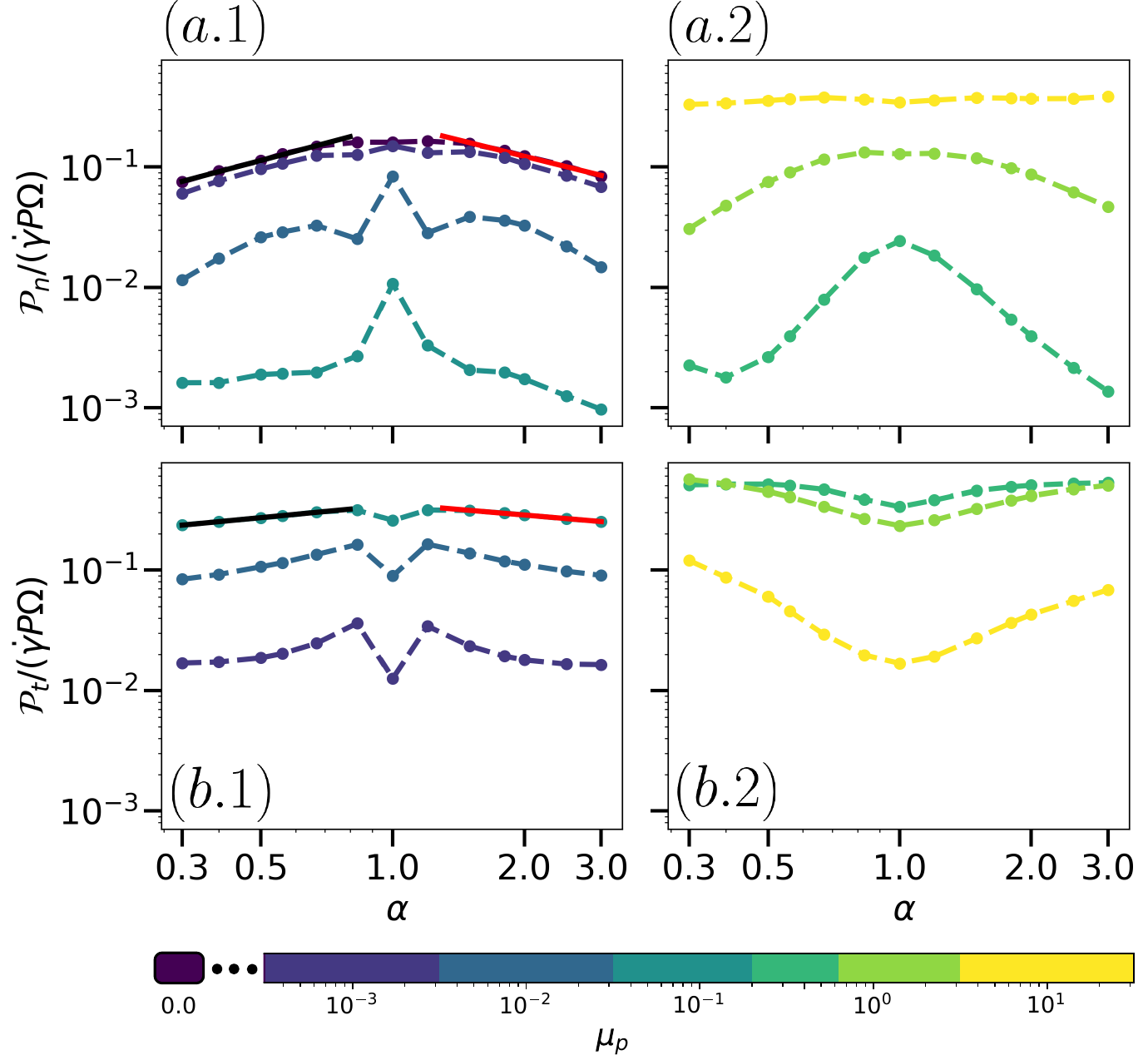}
    \caption{(a) Normalized normal and (b) tangential dissipation for $I\approx10^{-2}$ and different values of microscopic friction.
    On the left, data at $\mu_p<0.4$, on the right $\mu_p\geq0.4$. 
    In panel (a.1), the red (black) line represents the best-fit power law for frictionless prolate (oblate) particles, with a slope of $-0.90$, $(0.96)$.
    For particles with $\mu_p=0.1$, panel (b.1), the corresponding slopes are $-0.30$ for prolate and 
    $+0.33$ for oblate particles.
    From (b.1, b.2), it is evident that the tangential dissipation increases up to a value of $\mu_p$ between $0.1$ and $1.0$, depending on the shape, and then reduces for higher values.
    }
    \label{fig:dissipation_normal_tangential}
\end{figure}

As a function of the inertial number and the microscopic friction, the leading dissipative mechanism of granular flows and dense suspensions can be either due to normal inelasticity, as viscoelastic damping or to frictional sliding \cite{degiuli2016phase, trulsson2017effect, nagy2020flow}.  \\
We can split the dissipated power accordingly in a normal component $\mathcal{P}_{n}$, and a tangential one due to frictional sliding $\mathcal{P}_t$, in formulas:
\begin{align}
    \mathcal{P}_n = \sum_{i \in N_c} F_i^{n, \gamma} u_i^{n},   
    &&\mathcal{P}_t &= \sum_{i \in N_s} F_i^{t} u_i^{t} \, ,
\end{align}
where $F_i$ is the force due to contact $i$, $N_c$ is the total number of contacts, $u_i$ is the relative velocity of the grains at the contact point, $F^{n,\gamma}$ is the normal component of the force due to damping and $N_s$ is the number of sliding contacts, those which satisfy $F_i^t = \mu_p F_i^n$.
Figure \ref{fig:regime_diagram} shows for each aspect ratio the regime diagram of the dissipation as the level set curve where $\mathcal{P}_t=\mathcal{P}_n$.
The parametric space has three regions:
the normal dissipation regime for low $\mu_p$ and high $I$, the rolling or highly frictional regime for high $\mu_p$ and high $I$ and the frictional sliding regime for intermediate $\mu_p$ and low enough $I$.
For $\alpha>1$ the diagram closely resembles the one obtained for sphero-cylinders in \cite{nagy2020flow}. Here, we extend this diagram by also including $\alpha<1$.
For values of $|M_\alpha|>1.2$ the crossover from the normal dissipation to the sliding regime occurs at higher inertial number the higher the absolute value of the shape ratio (see the inset).
Regarding the boundary between the sliding and rolling regime, non-spherical particles start to roll at higher values of $I$ and $\mu_p$.\\
To rationalize our findings we visualize the power dissipation normalized by the injected power into the system, given by $\dot \gamma \sigma_{xy} \Omega \sim  \dot \gamma P \Omega$, in Figure \ref{fig:dissipation_normal_tangential}.
We observe that, up to moderate friction coefficients ($\mu_p \leq 0.4$), Figure \ref{fig:dissipation_normal_tangential}a, the power dissipated due to inelasticity decreases as particles become more elongated or flattened for aspect ratios $M_\alpha>1.2$.

\begin{figure}
    \centering
    \includegraphics[width=\linewidth]{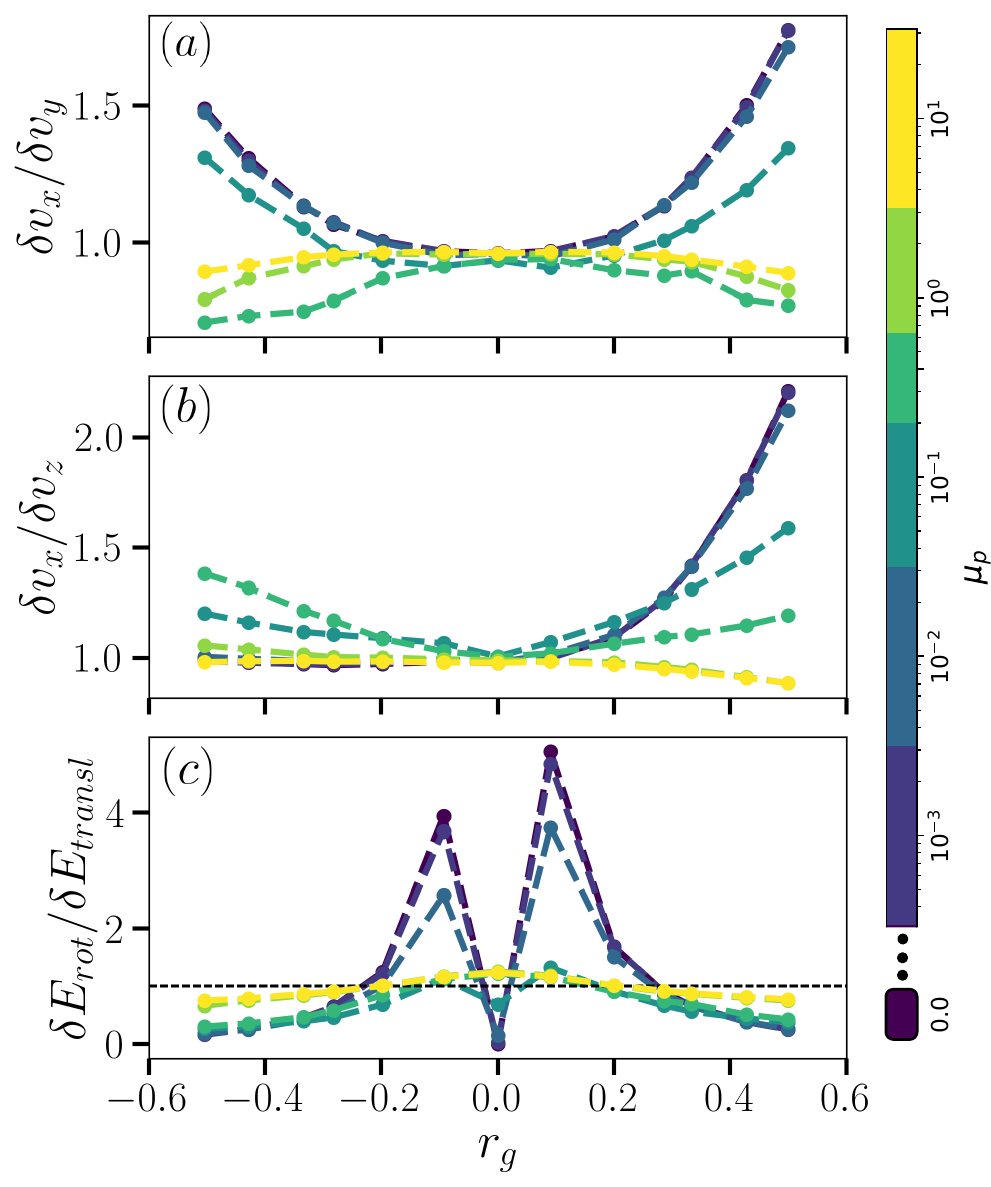}
    \caption{(a) Ratio of the streamline-wise velocity fluctuation $\delta v_x$ to the gradient-wise velocity fluctuation $\delta v_y$, (b) and to spanwise vorticity $\delta v_z$.
    (c) Ratio of rotational to translational kinetic energy, where the dashed line indicates a ratio of one.
    Different microscopic friction coefficients, with the same color legend as in Figure \ref{fig:dissipation_normal_tangential}.
    For low values of friction the flow component of the fluctuation dominates for prolate particles, whereas for oblate ones, both the flow and the vorticity ones are dominant. 
    In the sliding regime the fluctuations in the vorticity direction are reduced, as friction stabilizes that translation.
    For low values of friction and eccentricity the rotational fluctuations dominate.
    At high values of friction energy tends to be equally distributed between translation and rotation, as well as between the translational degrees of freedom.
    Data taken at $I\approx 0.01$.}
    \label{fig:v_fluct_ratio}
\end{figure}

Owing to the system's structural anisotropy, assuming all particles are perfectly aligned with their major axis along the flow direction, the cross-sectional areas of each grain are listed in Table \ref{tab:cross_sections}.  
\begin{table}
\caption{Normalized cross-sectional areas and area ratios for oblate and prolate ellipsoids, with major axis aligned with the streamlines.}
\label{tab:cross_sections}
\begin{tabular}{c|c|c|c|c|c}
 & $A_{yz}/\pi a^2$ & $A_{xz}/\pi a^2$ & $A_{xy}/\pi a^2$ & $A_{yz}/A_{xz}$ & $A_{yz}/A_{xy}$ \\
\hline
Oblate  & $ \alpha$     & $1$           & $\alpha$     & $\alpha$           & $1$ \\
Prolate & $1$            & $ \alpha$    & $ \alpha$     & $1/\alpha$         & $1/\alpha$
\end{tabular}
\end{table}
Among them, the gradient-vorticity cross-section $A_{yz}$ is the smallest, implying that collisions are least likely in the flow direction.
Defining the particles' velocity fluctuation as the excess over the linear velocity profile $\delta \boldsymbol{v}= \boldsymbol{v} - \dot \gamma y \mathbf{e}_x$, velocity fluctuations along the flow direction, $\delta v_x$,  tend to persist longer, due to the reduced cross-sectional area.
Moreover, as reported in \cite{guo2013granular} at high volume fractions and seen in Figure \ref{fig:v_fluct_ratio}a and b, the dominant fluctuation is $\delta v_x$, indicating that most of the fluctuating kinetic energy is in the flow direction.
Taken together, these effects offer a simple and consistent explanation for the reduced normal power dissipation observed in systems with anisotropic particles: lower collision frequency along the dominant fluctuation direction leads to less energy lost through collisions.
This simple geometrical argument also clarifies why, for flattened particles, fluctuations in the vorticity direction are of the same magnitude as those in the flow direction, as their respective cross-sections are equal, $A_{xz}=A_{xy}$.\\
From Figure \ref{fig:dissipation_normal_tangential}, we also observe that in the frictionless case $\mathcal{P}_n \sim M_\alpha^{-1}$.
In the limit $\mu_p \to +\infty$, the influence of shape on the normal component of dissipation, which is the dominant one in this regime, becomes negligible, and all velocity fluctuations tend to attain similar values, see Figure \ref{fig:v_fluct_ratio}a and b.
In this limit alignment is reduced, and the effective cross-sectional areas approach those of the volume-equivalent spheres.

In the sliding regime, the influence of particle shape on the frictional dissipation is less marked, with non-spherical particles dissipating virtually all the injected power to overcome friction.\\
Previous scaling arguments for circular particles estimated that the cross-over from normal dissipation to sliding regime occurs at $I_c \sim \mu_p^2$ \cite{degiuli2016phase}, as indicated by the dashed line in Figure \ref{fig:regime_diagram}.
In our case, we find roughly $I_c \sim f(\alpha) \mu_p^2$, where $f(\alpha)$ a non-monotonic function, with global minimum in $\alpha=1.0$.
The reduced normal dissipation of aligned particles, shifts the transition to higher $I_c$ for both elongated \cite{nagy2020flow} and flattened particles.
For the latter, their slightly higher alignment shifts the transition at slightly higher inertial number compared to the respective reciprocal shapes.\\
From Figure \ref{fig:regime_diagram}, we also observe that at a fixed inertial number $I$, the transition from rolling to sliding regime occurs at higher values of $\mu_p$ for nonspherical particles.
Oblate shapes start rolling at higher $\mu_p$ than prolate ones with similar eccentricity, as they experience more anisotropic fabric and comparatively stronger forces (Sections \ref{sec:global_force}, \ref{sec:particle_data}).
Interestingly, this second transition from sliding to rolling is monotonic across all $M_\alpha$, indicating that in the rolling regime, lower volume fraction and less collective motion reduce geometric constraints; see Sections \ref{sec:rheology} and \ref{sec_correlation_length}. 

\begin{figure}
    \centering
    \includegraphics[width=\linewidth]{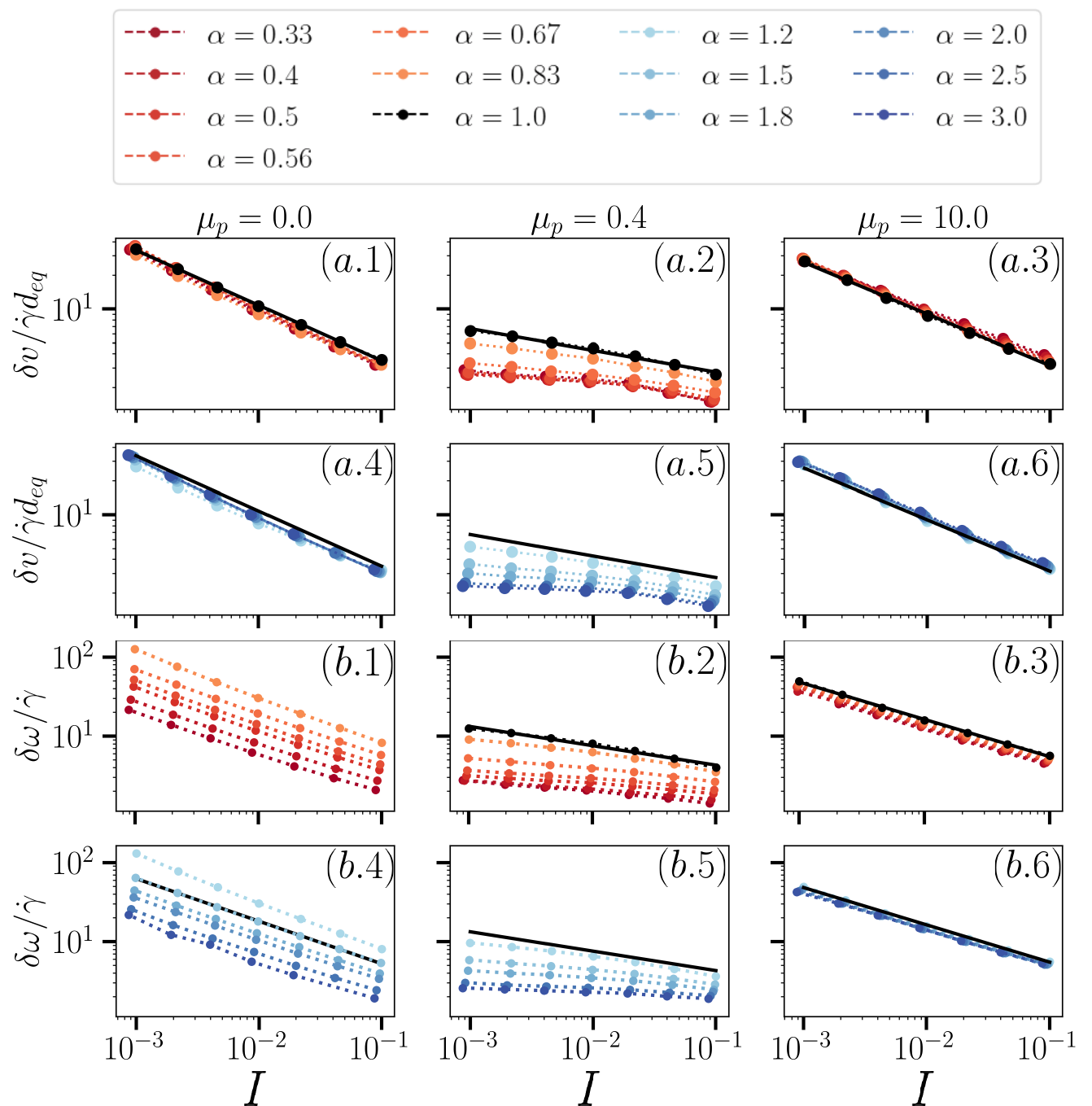}
    \caption{
    (a.1-6) Intensity of normalized velocity fluctuations, \( \delta v = |\delta \boldsymbol{v}| \), and (b.1-6) angular velocity fluctuations, \( \delta \omega = |\delta \boldsymbol{\omega}| \).  
    Results are shown for \( \mu_p = 0 \) (left), \( 0.4 \) (center), and \( 10.0 \) (right).  
    Black lines indicate best-fit power laws for spherical particles, with slopes of \(-0.49\), \(-0.19\), and \(-0.46\) (a.1-6), and \(-0.54\), \(-0.25\), and \(-0.47\) (b.1-6), where in the frictionless case the fit is performed on $\alpha=1.5$.
    Splitting the data for elongated and flattened particles makes the symmetry of reciprocal shapes less evident, but clearly shows the monotonic trend. 
    }
    \label{fig:vel_fluctuations}
\end{figure}

Next, we turn our attention to the non-monotonic transition from normal dissipation to sliding regime of particles of small elongation, i.e. $M_\alpha \leq 1.2$, in the normal dissipation regime.
Previous studies showed how even a small degree of eccentricity, dramatically changes the microscopic dynamics both in static and dynamic conditions \cite{keta2020translational, sun2024evident}. 
Firstly, we recall that in all the parametric space, they mostly reach a weakly nematic phase (Figure \ref{fig:S2_theta_omega_z}a).
Therefore, for these grains the caging effect is smaller and their rotation is less frustrated.
We measure the fluctuating kinetic energy, that we decompose in the translational component $\delta E_{transl} =
0.5\sum_{i=1}^N m_i ||\delta \boldsymbol{v}_i||^2$ and the rotational one $\delta E_{rot} = 0.5\sum_{i=1}^N  \delta \boldsymbol{\omega}_i \cdot \mathbf{J}_i \delta \boldsymbol{\omega}_i$, where $\mathbf{J}_i$ is the moment of inertia tensor of the $i$-th particle, and $\delta \boldsymbol{\omega} $ its fluctuating angular velocity, measured as $\delta \boldsymbol{\omega}_i = \boldsymbol{\omega}_i - \langle \boldsymbol{\omega}\rangle$. 
Figure \ref{fig:v_fluct_ratio}c shows that for low friction, $\mu_p<0.01$, the fluctuating rotational energy exceeds the translational one.
Therefore, the previous scaling arguments \cite{degiuli2016phase, trulsson2017effect} exclusively based on the translational fluctuating energy need to be revisited to determine $\mathcal{P}_n$ in the case of slightly aspherical particles.
One does also notice how for highly eccentric particles the rotational component is considerably smaller for $\mu\leq0.4$, which makes the scaling arguments in \cite{degiuli2016phase} still applicable even for non-spherical particles.
This motivates why the normal dissipation to sliding transition lines have approximately the same slope, recall Figure \ref{fig:regime_diagram}. 
Notably, for high microscopic friction the two components of the fluctuating kinetic energy tend to attain a similar magnitude, indicating a thermalizing effect, which is also reflected in the velocity fluctuations.

In Figure \ref{fig:vel_fluctuations} we report the intensity of velocity fluctuations in the different regimes.
Defining $\mathcal{L} = \delta v / \dot{\gamma} d_{eq}$, the so-called lever effect, which measures the dimensionless intensity of the velocity fluctuations ($\delta v$), in the normal dissipation and rolling regimes, $\mathcal{L} \sim I^{-\xi}$, with $\xi \approx 0.5$ as reported by multiple studies \cite{peyneau2008frictionless, degiuli2016phase, trulsson2017effect, kharel2017vortices, rognon2021shear}.
In the sliding regime however, we find $\mathcal{L} \sim I^{-0.19}$ for spherical particles, slightly lower than the one reported in 2D \cite{degiuli2016phase}.
For elongated particles, the fluctuations are reduced even more in the sliding regime, confirming the more coherent flow they undergo.
In the frictional-rolling regime the shape of the particles has very little influence on the velocity fluctuations, signaling that all particles behave in a similar way as alignment is reduced, and the fluctuations tend to become isotropic, recall Figure \ref{fig:v_fluct_ratio}.

Looking at the fluctuations of angular velocity, in the frictionless and highly frictional case we observe $\delta \omega \sim I^{-1/2}$, analogous to the velocity fluctuations in the same regimes.
In the frictionless limit, albeit following the same same power-law scaling, the angular velocity fluctuations do not collapse onto a single curve across different particle shapes. 
Instead, the degree of fluctuation decreases monotonically with increasing particle eccentricity, indicating that more aspherical particles suppress rotational fluctuations even in the absence of friction.
This is in contrast with what happens for the translational fluctuations, in this regime, which collapse regardless of the shape and also to $\delta \omega$ in the rolling regime, where particle shape does not play a big role.
In the sliding regime, the angular velocity fluctuations' intensity is reduced and the trend with respect to $I$ is less pronounced, again comparable to the velocity fluctuations.

\subsection{Rheological properties}
\label{sec:rheology}

\begin{figure}
    \centering
    \includegraphics[width=\linewidth]{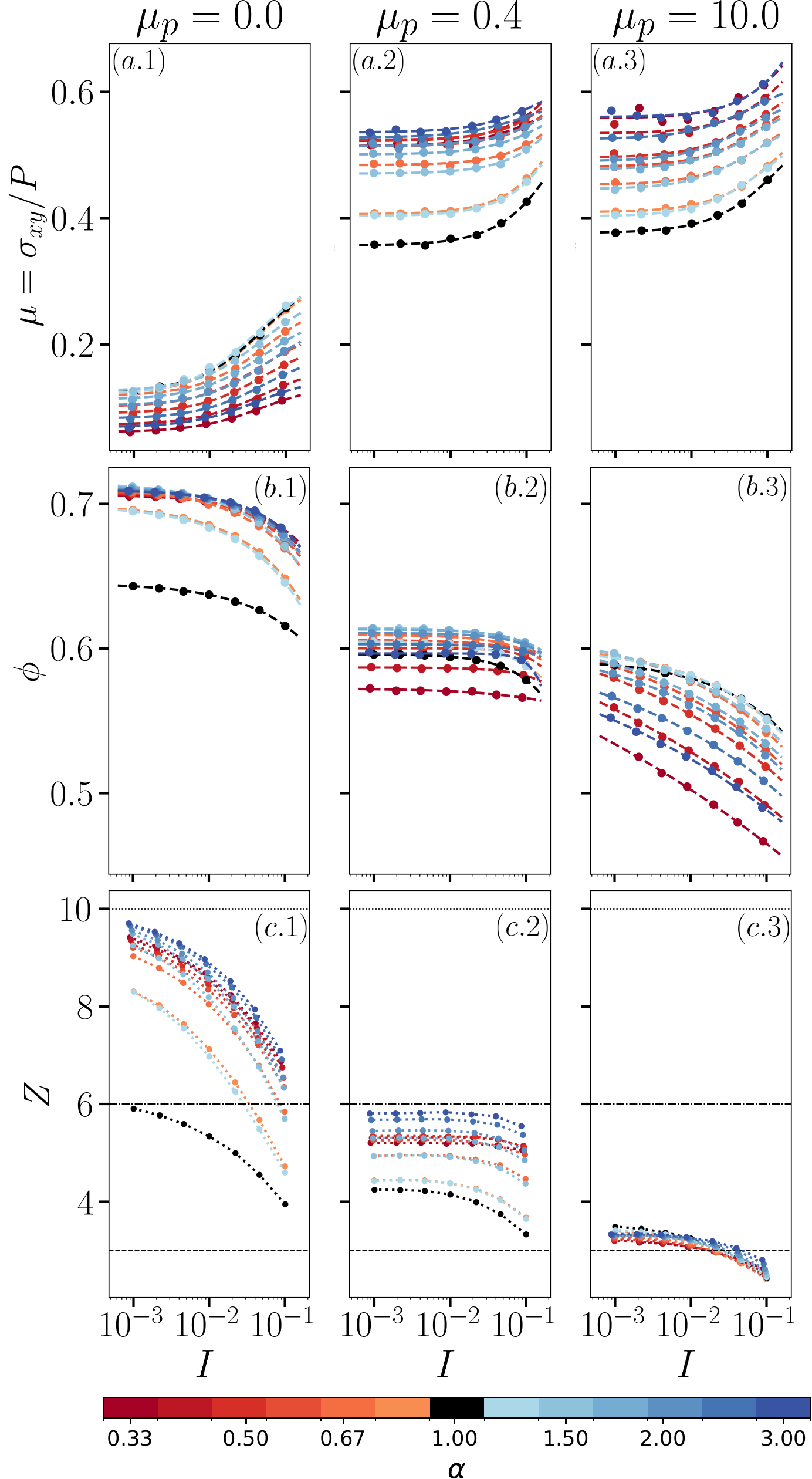}
    \caption{Stress anisotropy (top) and packing fraction (middle), and average number of contacts per particle (bottom) for different values of microscopic friction. 
    The dashed, dotted-dashed and dotted lines indicate $Z=3, 6, 10$ respectively.}
    \label{fig:mu_I}
\end{figure}
\begin{figure}
    \centering
    \includegraphics[width=\linewidth]{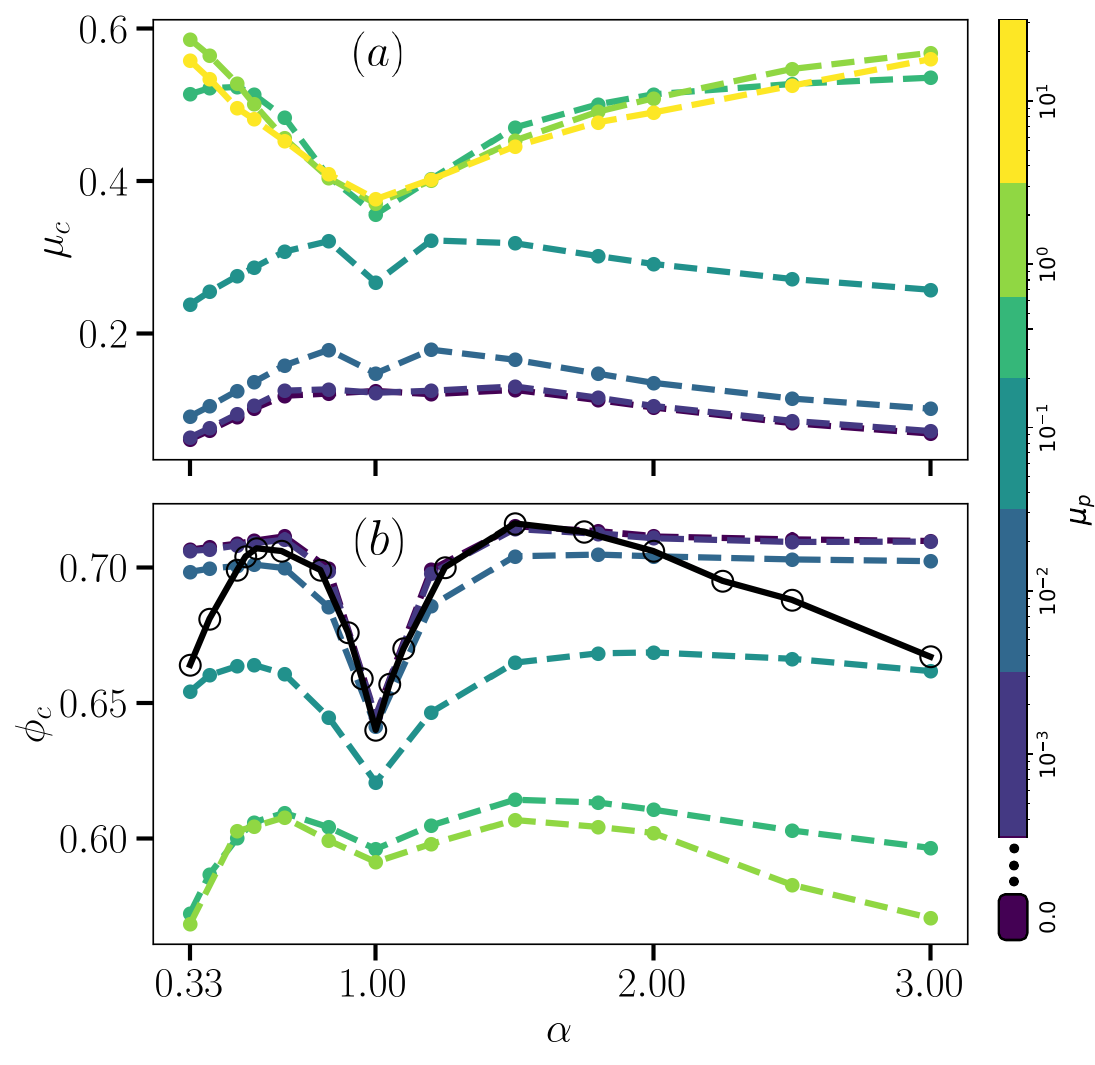}
    \caption{Critical shear strength (top) and packing fraction (bottom) as a function of aspect ratio.
    For $\phi_c$ values of \textit{random close packing} from \cite{donev2004improving} are also reported with a black line and open circles, while values for $\mu_p=10.0$ are not reported since simulations at the lower values of $I$ are needed to get a meaningful estimate.
    }
    \label{fig:mu_and_phi_c}
\end{figure}

Having characterized the fluctuations of the grains, we move on to their rheology.
In this section we present the best fit of effective friction and packing fraction with respect to the standard $\mu(I)$ rheology \cite{gdr2004dense}, $\mu(I)=\sigma_{xy}/P$, for different miscroscopic friction, see Figure \ref{fig:mu_I}a and b.
While more fundamental kinetic theories exist, this phenomenological model is chosen for its proven effectiveness in capturing the essential flow behavior and its role as a standard benchmark, allowing for direct comparison with other studies in the field.
The well-known formulas are:
\begin{align}
    \mu(I) = \mu_c + {c_\mu \over 1 + I_0/I} \\
    \phi(I) = \phi_c - c_\phi I^{\beta_\phi} \,,
\end{align}
where $\mu_c, \, \phi_c$ are the critical effective friction and packing fraction respectively, while $I_0, c_\mu, c_\phi$ are constants and $\beta_\phi$ is an exponent.
Similarly to \cite{trulsson2018rheology,nagy2020flow} the effective friction generally increases by increasing particle elongation (or flatness) and microscopic friction.
The actual values for $\alpha=2$ are almost identical to those measured by \cite{nagy2020flow} for spherocylinders with the same aspect ratio, suggesting that $\alpha$ is a good parameter to characterize the shape in granular rheology, beyond the characteristic size of the grain.
See supplemetary Figure \cite{bilottosupp} \cite{bilottosupp} for a direct comparison between spheroids and data for sphero-cylinders taken from \cite{nagy2020flow}.
\\
We observe that it is slightly easier to shear flat particles for the frictionless case with respect to elongated ones with the same value of $|r_g|$, which is also reflected by the fact that oblate spheroids attain a higher alignment, which minimizes $A_{yz}$. 
For $\mu_p=0.4$ slightly oblate particles, $0.66\leq \alpha \leq 1.0$, are harder to shear, but the opposite is true for more elongated ones.
As $\mu_p$ increases, deviations from the spherical shape make shearing harder, since the alignment that made sliding efficient without friction, now comes at a higher cost.
Nevertheless as the mechanism of dissipation reverts back to normal dissipation at very high microscopic friction, the effective friction of the granular assembly keeps a finite values, even in the limit $\mu_p \to \infty$.\\
The critical shear strength at shear jamming, $I\to 0$ $\mu_c$ is shown in Figure \ref{fig:mu_and_phi_c}a.
As a general trend at a fixed $\alpha$ increasing microscopic friction increases the critical strength, however, in the highly frictional case $\mu_p=10$ highly non-spherical particles have a lower critical shear strength compared to $\mu_p = 1.0$.
This apparent non-linearity is due to the wider distribution of particle orientations in the former case, with misaligned particles having more space to move without sliding.

Similarly to the shear strength, the effect of friction on the packing fraction is highly nonlinear, see Figures \ref{fig:mu_I}b and \ref{fig:mu_and_phi_c}b. 
In the low friction limit, more elongated or flat particles reach higher densities, whereas these trends become non-monotonic as friction comes into play.
This behaviour is similar to what is observed for random packings of non-spherical particles \cite{donev2004improving, wouterse2007effect}, with $\mu_p$ providing the randomizing effect.
Interestingly, the values of $\phi_c$ for low friction and small eccentricity lie close to those measured by \cite{donev2004improving} and defined as random close packings.
However the effect of shear becomes prominent at high degree of eccentricity where grains can achieve significantly higher density than if randomly packed.

Figure \ref{fig:mu_I}c shows the average number of contacts per particle as a function of the inertial number. 
As expected, an increase in the inertial number leads to a decrease in the average coordination number, reflecting the transition from quasi-static, solid-like behavior to a dynamic, fluid-like regime.
This decrease is more pronounced in the frictionless case, where particles rely solely on geometric hindrance for stability and cannot resist tangential forces that would otherwise help maintain contact.
A clear dependence on particle shape is also evident: both frictionless and frictional particles with $\alpha<0.5$ exhibit significantly fewer contacts than those with $\alpha>2.0$.
In all cases, the average coordination number remains below the isostatic limit — 6 for frictionless spheres and 10 for frictionless ellipsoids — indicating a hypostatic state.
The deviation from isostaticity is particularly strong for oblate particles, which can be attributed to their flatter geometry.
As noted in \cite{donev2007underconstrained}, contacts on more rounded surfaces, \emph{i.e.} lower curvature, stabilize rotational degrees of freedom more easily.\\
For high friction, aspherical particles achieve a lower volumetric density and their coordination becomes similar to that of spheres.\\
The values measured in Figure 7b and c are in most cases slightly higher than the critical values for the appearance of elastic stresses, proportional to $E$, as measured under constant volume conditions in \cite{berziDenseShearingFlows2022}, provided a direct comparison between cylinders and spheroids of the same aspect ratio is appropriate.
Additional evidence is found on the extended plateaus of $\phi$, $Z$ with respect to inertial number in Figure 7b.2 and 7c.2, which signal a quasi-static (elastic) response \cite{campbellStresscontrolledElasticGranular2005}.
Additional results on normal stress differences can be found in Appendix \ref{app:normal_stresses}.

\subsection{Correlation of velocity fluctuations}
\label{sec_correlation_length}

The emergence of length scales of the velocity fluctuations, gives us a measure of how correlated the motion of grains is.
Here we measure the two-point autocorrelation function of the vertical velocity fluctuations, computed with the following formula:
\begin{align}
    \tilde{C}(|\mathbf{r}|) = { \langle \delta v_y(\mathbf{x}+\mathbf{r}) \delta v_y(\mathbf{x}) \rangle 
    \over 
    \langle \delta v_y^2(\mathbf{x})\rangle} \,
\end{align}
where the average is taken over all points $\mathbf{x}$ and then over all time steps.

\begin{figure}
    \centering
    \includegraphics[width=.9\linewidth]{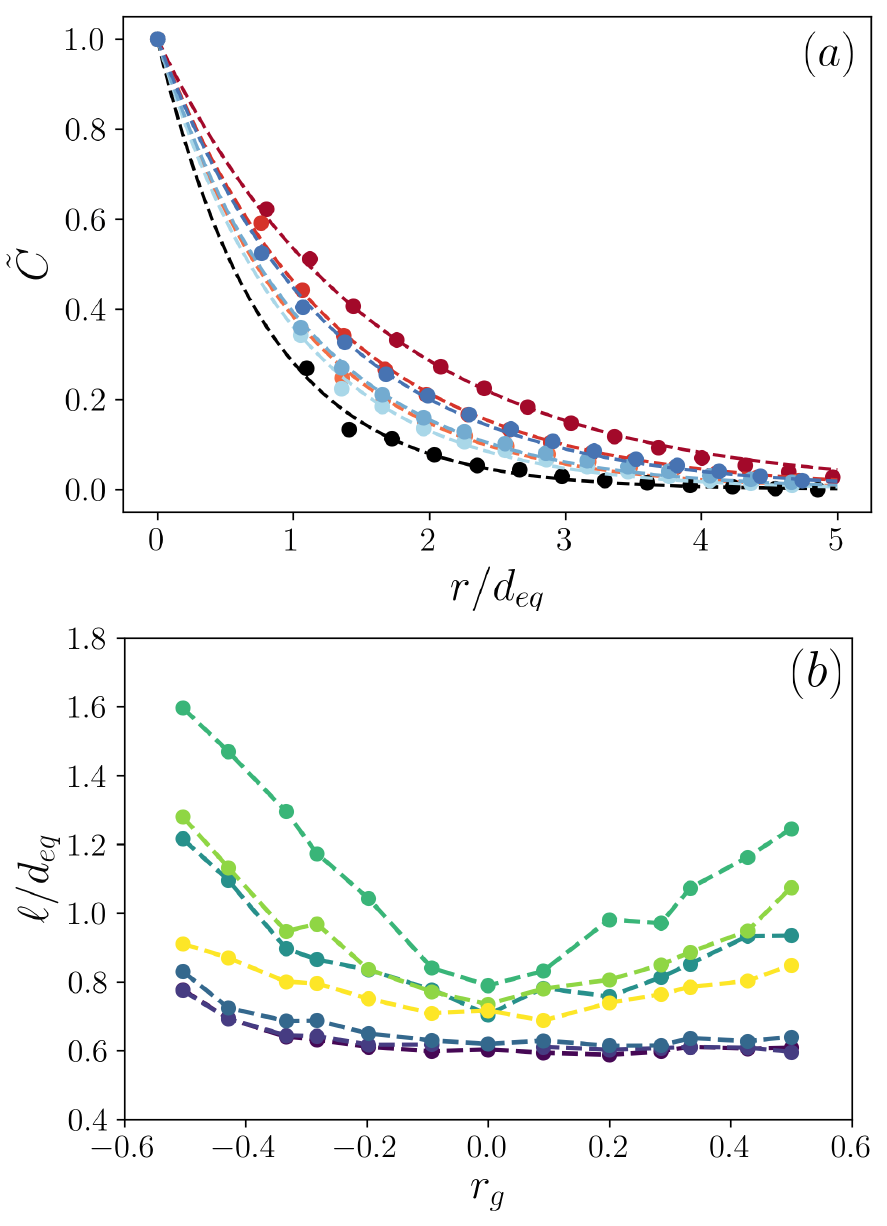}
    \caption{(a) Spatial velocity autocorrelation function, computed for $I=10^{-1}$, $\mu_p=0.4$ (colors as in Figure \ref{fig:mu_I}).
    For all grain shapes the function is well fitted by an exponential decay, dashed line.
    Non-spherical particles, display higher spatial correlation than spheres.
    (b) Correlation length as a function of shape ratio at fixed $I=10^{-1}$ for different microscopic frictions (colors as in Figure \ref{fig:mu_and_phi_c}).
    The correlation length is higher for microscopic friction $0.1\leq \mu_p \leq 1.0$, \emph{i.e.} in the frictional sliding regime, and grows as a function of the shape ratio.
    }
    \label{fig:correlation} 
\end{figure}

We observe that for all cases the autocorrelation is well fitted by an exponential decay $\tilde{C}(r) \approx e^{-r \over \ell}$ see figure \ref{fig:correlation}a.
Next, we plot the correlation length as a function of the aspect ratio for different microscopic friction coefficients, figure \ref{fig:correlation}b.
The sliding regime shows the highest correlation lengths, with elongated and flattened particles motion being much more entangled.
Specifically, oblate particles correlate with longer lengths, which is an explanation why the sliding regime extends further for lentil-like particles.

A simple estimate of the probability of a particle collision during a vertical displacement $\delta y$ can be obtained from the product of the displacement, the effective cross-section, $A_{xz}$, and the number density of particles ($n  = \phi /V_p$). 
If the spheroids align with their major axis along streamlines, we approximate the collision probability as:
\begin{align}
    p_{coll} \sim n \delta y A_{xz} = \begin{cases}
        \phi \delta y /(\alpha a),\, &\text{oblate} \\
        \phi \delta y/a, \, &\text{prolate}
    \end{cases}
\end{align}
Therefore, provided the variations in volume fraction between particles are small, oblate spheroids are more likely to collide than prolate ones. 
At high densities, and with some degree of energy dissipation, particles' motion tends to become correlated due to collisions.
This simple argument can motivate the behavior in the frictionless case, see purple curves in Figure \ref{fig:correlation}b.
However, this line of reasoning falls short when considering frictional particles. 
Notably, frictional prolate spheroids display significantly longer-ranged correlations in their motion compared to frictional spheres, even when the packing densities are similar.
This suggests that factors beyond simple packing geometry must be at play.\\
A likely explanation lies in the enhanced coupling between rotational and translational degrees of freedom introduced by both particle shape and friction \cite{zou2022microscopic}.
Non-spherical spheroids experience more anisotropic contact distributions, see also Figure \ref{fig:fabric_anisotropy}.
When friction is present, these elongated particles can transmit torques more effectively through their contacts, linking rotations and translations across larger regions of the system. 
This interdependence facilitates coordinated clusters of motion, where the reorientation of one particle affects not only its immediate neighbors but also more distant particles through a network of frictional constraints \cite{guo2013granular}.\\
In simple terms, friction can lock particles into local configurations for longer periods, promoting the build-up of correlated structures that persist over time.
The resulting dynamics are no longer dominated by binary collisions but instead reflect the collective motion of strongly interacting clusters, particularly in the case of anisotropic shapes, see Appendix \ref{app:diffusion} for more detail.\\
As stated earlier, the higher correlation of motion for flattened particles, is another factor leading to the extended sliding regime they display compared to elongated ones, since to break the correlation higher values of $\mu_p$ or $I$ are needed.
To better characterize the correlation length, additional simulations with larger system sizes are required to exclude finite size effects.
Notwithstanding, when particles are allowed to overlap in 3D the correlation length is expected to increase up to the volume fraction where shear jamming is reached \cite{berziSteadyShearingFlows2015, oyamaAvalancheInterpretationPowerLaw2019}.
As previously mentioned, in several of the simulations, especially those in the sliding regime at low $I$, we are close to this limit.
The values reported here are reasonably smaller than the system size, with the maximum ratio of $\ell/L_y=0.16$ measured at $I=0.1$, and the trend shown is consistent along all inertial numbers object of this study: although not quantitatively rigorous, the qualitative trends observed are deemed trustworthy.

\subsection{Contact fabric in the sliding regime}
\label{sec:global_force}
Here and in the next section our analysis focuses on the sliding regime at $\mu_p=0.4$.
To characterize the contact distribution in space, we compute the contact normals fabric tensor.
Denoting $\mathbf{n^c}$ the contact normal, the discrete version of the fabric tensor is:
\begin{align}
    \mathcal{F}_{ij} = {1 \over N_c} \sum_{k=1}^{N_c} n^{c, k}_i n^{c, k}_j \, ,
\end{align}
By definition, the fabric tensor is symmetric, thus it can be diagonalized by changing basis.
The eigenvalues, which verify the normalization condition $\lambda_1+\lambda_2+\lambda_3=1$, reflect the number of contacts along the preferential directions.
Regardless of the parameters, there is always an eigenvector aligned with the vorticity direction.
\begin{figure}
    \centering
\includegraphics[width=0.95\linewidth]{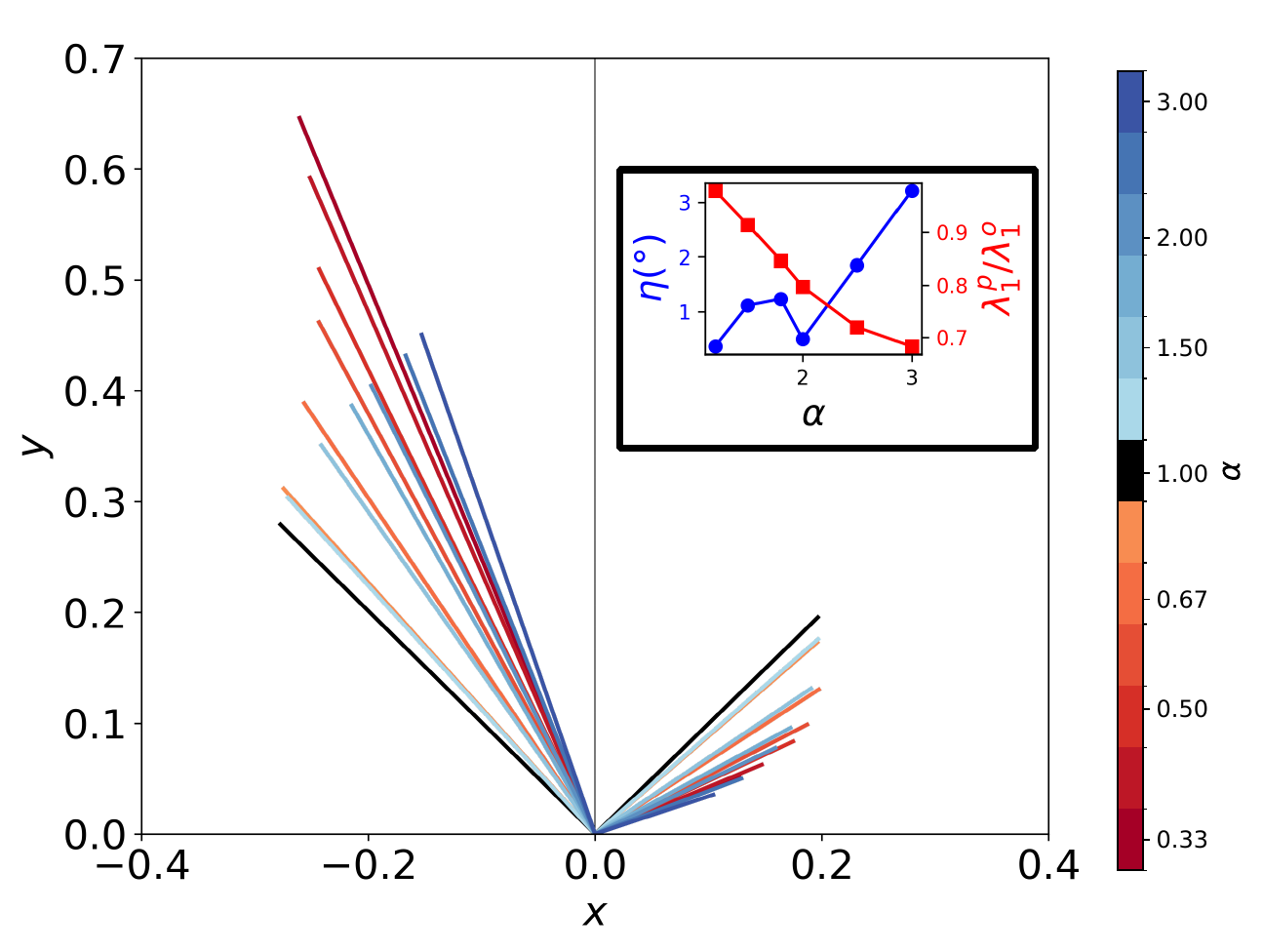}
    \caption{Fabric eigenvectors scaled by their corresponding eigenvalue at $I\approx 0.01$. 
    In the inset: the angle $\eta$ (in blue) between the largest eigenvectors of the reciprocal shapes, \emph{e.g.} $\alpha_1=3, \, \alpha_2=1/\alpha_1=1/3$, the ratio between the biggest eigenvalue of the prolate spheroid with respect to the reciprocal oblate one, $\lambda_1^p/\lambda_1^o$, in red.
    }
    \label{fig:fabric_anisotropy}
\end{figure}
In Figure \ref{fig:fabric_anisotropy} we report the two eigenvectors in the $xy$ plane scaled by their respective eigenvalue.
For spheres the eigenvectors are roughly oriented at $\pm \pi/4$, in the extension and compression direction of the flow respectively.
As $|r_g|$ increases, the eigenvectors rotate clockwise similarly to the force distribution, see Section \ref{app:global_contact_distribution}.
The eigenvalue in the compression direction is greater than in the extension one in the spherical case. 
The anisotropy, measured as the ratio of the two biggest eigenvalues, increases with eccentricity.
Additionally, one notices that the eigenvalues in both the compression and extension direction of oblate particles are greater than the one for prolate ones, 
suggesting that fewer contacts take place in the vorticity direction for lentil-like particles.
The full planar distribution of contacts can be found in Appendix \ref{app:global_contact_distribution}.
As shown in the inset of Figure \ref{fig:fabric_anisotropy}, the anisotropy of oblate particle is higher than the corresponding reciprocal prolate ones.
Despite alignment of reciprocal shapes being similar, recall Figure \ref{fig:S2_theta_omega_z}a, the much more anisotropic fabric makes this alignment harder to disrupt, hence the transition to the rolling regime happens for higher values of $\mu_p$.
The angle of the principal eigenvectors shows only a modest discrepancy.

\subsection{Particle contact data in the sliding regime}
\label{sec:particle_data}

To transition from spatial interaction data to local, particle-level information, we bin contact data on the surface of ellipsoids according to the polar angle between each contact point and the ellipsoid's axis of symmetry ($\psi\in [-\pi/2, \pi/2]$).
Due to symmetry we divide the surface in 10 sections ($\psi\in (0, \pi/2)$) with constant intervals of $\psi$ (and not constant surface area).
Let us define the surface density of contacts $\rho_s$ so that 
\begin{align}
    \int_A \rho_s dA = 1 \, .
\end{align}
\begin{figure}
    \centering
    \includegraphics[width=\linewidth]{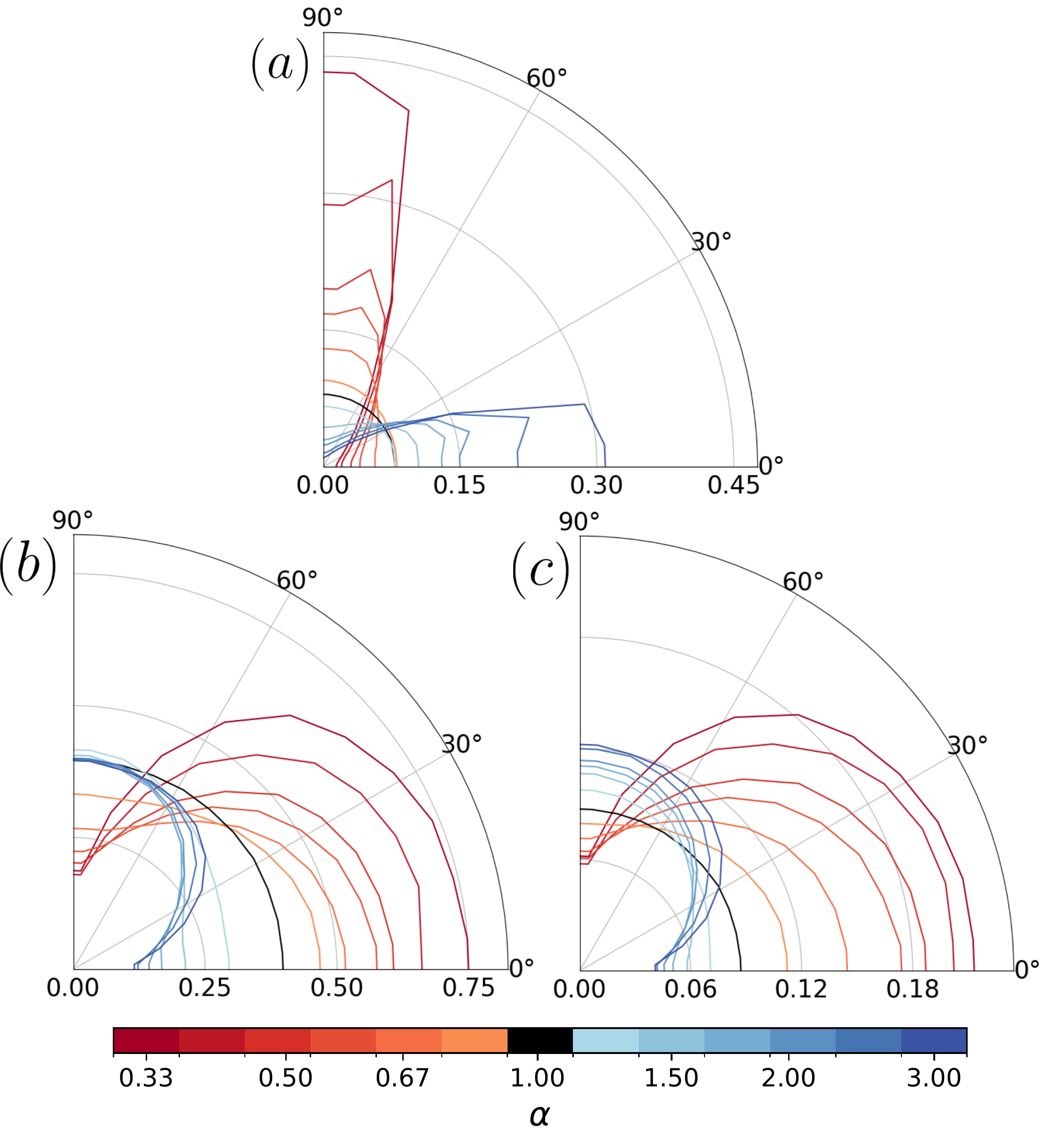}
    \caption{Polar histograms for contact data in the ellipsoid reference frame.
    The angle is measured from the ellipsoid symmetry axis.
    (a) Surface probability distribution of contacts, $\rho_s$.
    (b, c) Average normal and tangential force conditional to a contact being present normalized by average pressure times the surface area of the particle, $\overline{F}_n/PA_p$, and $\overline{F}_t/PA_p$ respectively. 
    It appears evident that the contact density and the force intensity are anticorrelated.
    }
\label{fig:local_contact_data}
\end{figure}
We also compute the average force conditional to a contact being present, both in the normal $\overline{F}_{n}(\psi)$ and in the tangential $\overline{F}_{t}(\psi)$ direction.
To get the average force depending on the polar angle we can simply compute $\langle F(\psi) \rangle = \pi a \sin{\psi} \sqrt{a^2\cos^2{\psi}+c^2\sin^2{\psi}} \rho_s(\psi)\overline{F}(\psi)$.
The separate computation of the two makes the intensity of the local interactions evident.

\begin{figure*}
    \centering
    \includegraphics[width=\linewidth]{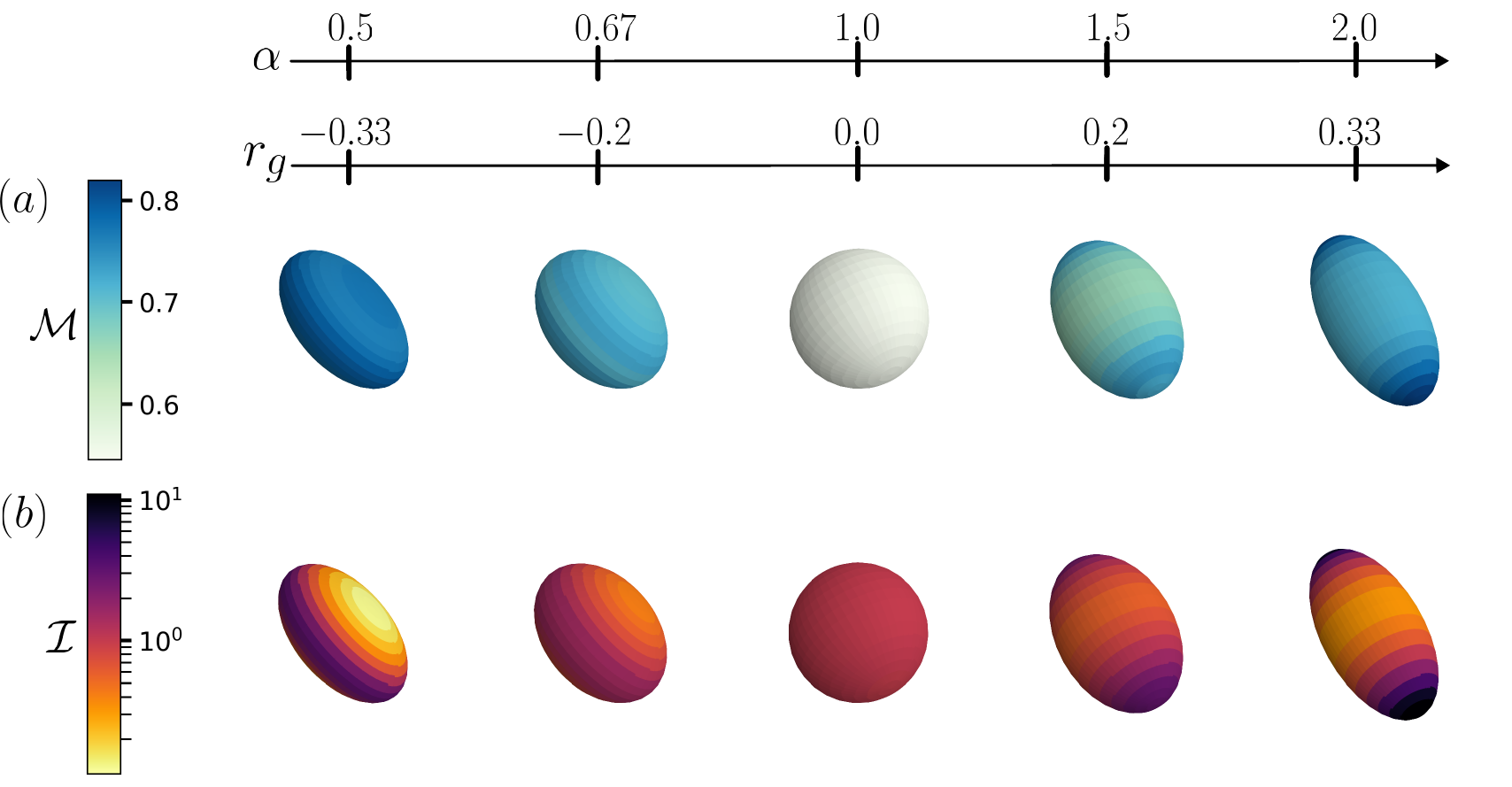}
    \caption{(a) Average surface distribution of friction mobilization, $\mathcal{M} = F_t/\mu_p F_n$, for different particle shapes at $I\approx 0.01$, $\mu_p=0.4$.
    Non-spherical particles, mobilize significantly more friction than spheres, with higher mobilization on the major axis/axes.
    (b) Surface power dissipation density inhomogeneity,  $\mathcal{I}$, for the same values of inertial number and microscopic friction coefficient.
    A value smaller than one indicates less dissipation compared to the homogeneous case, whereas a value greater than one represents a hot spot for dissipation.
    The color-map is in logarithmic scale.}
    \label{fig:surface_ellipsoid}
\end{figure*}

As reported in Figure \ref{fig:local_contact_data}a, the probability of contact is uniform $\rho_s = 1/ 4 \pi a^2$ for spherical particles .
For elongated and flattened particles, contacts per unit area are more likely to occur on the major axis that is tilted in the flow direction. 
However, most of these contacts are weak, Figure \ref{fig:local_contact_data}b and c, as shown by the distributions of forces, that have the maximum on the minor axis.
Therefore, while non-spherical particles have on average more contacts (see Figure \ref{fig:mu_I}), they have a few strong contacts acting on the minor axis and several weak contacts on the major one \cite{azema2012force, zou2022microscopic}.
Following the same line of reasoning as \cite{nagy2020flow}, one can show that for a spheroid, the only points where an applied force causes no rotation are the two at $\psi = \pm \pi/2 $, and those at $\psi = 0$.
However, contrary to the case of sphero-cylinders all those points are mechanically unstable, even though those along the equator minimize the arm of the force with respect to the centroid. \\
Next, we look at the local friction mobilization defined as the ratio $\mathcal{M}= F_t/\mu_pF_n$. 
A value of $\mathcal{M}=0$ means that all tangential springs are inactive, whereas a value of $\mathcal{M}=1$ indicates that all contacts have reached the sliding limit. 
In Figure \ref{fig:surface_ellipsoid}a we report data for $\mu_p= 0.4$ and different degrees of elongation.
As found in previous studies \cite{azema2012force, zou2023particle}, elongated particles mobilize more friction compared to spherical ones. 
The higher mobilization is closely related to the extended frictional sliding regime shown in Figure \ref{fig:regime_diagram}.
The cause is that spherical particles can more easily rotate when a tangential force is applied, contrary to non-spherical ones (see Section \ref{sec:alignment}.
Interestingly, for low degrees of eccentricity oblate particles mobilize more friction than prolate ones, while for higher eccentricity the converse is true.
This nonlinearity reflects the same nonlinearity observed for the effective friction $\mu(I)$, due to the ordering.\\
Defining $\mathcal{P}_j$ as the average power dissipated on the surface
area element $A_j$ conditional to a contact being present,
and $\mathcal{P}_p$ as the total average power dissipated on the particle surface, we compute the spatial inhomogeneity of power dissipation density as $\mathcal{I} = \mathcal{P}_j \rho_s A_p^2 / (\mathcal{P}_{tot} A_j)$.
A value of $\mathcal{I}=1$ indicates a homogeneous dissipation, while values $>1$ and $<1$ indicate hot spots and cold spots, respectively.
It can be observed, see Figure \ref{fig:surface_ellipsoid}b, that the local power dissipation on the surface of the particles clusters in the same locations, where slip events are more common.
For high degrees of eccentricity, more than ten times dissipation occurs on the major axis compared to the minor one.
Accordingly, we propose the following microscopic mechanism: non-spherical particles establish a small number of strong contacts along their minor axis, where sliding is rare. To accommodate the linear velocity gradient, the particles exhibit a wobbling motion, resulting in the majority of power dissipation occurring at weaker contacts along the major axis.
A more detailed microscopic analysis of the contribution of frictional forces to shear strength can be found in Appendix \ref{app:stress_decomposition}.\\

\section{Summary and outlook}
\label{sec:final}
In this study, we have numerically investigated the rheological properties of granular shear flows composed of non-spherical particles by analyzing their dissipation mechanisms, velocity fluctuations, and stress distributions. 
Our findings highlight the significant influence of fully three-dimensional particle shape, even in quasi-two-dimensional flows.

We found that non-spherical particles generally tend to align with the flow, leading to a reduction in angular velocity and its fluctuations—except in the highly frictional rolling regime, where particles tend to follow the flow vorticity and the fluctuations collapse onto a simple power law. Interestingly, oblate particles rotate less individually and display more correlated motion, a trend we qualitatively explained using simple geometrical arguments.\\
Our dissipation regime analysis revealed that flat, lentil-like particles exhibit an extended sliding regime compared to elongated, rice-like particles, even when the shape ratio is kept constant. This behavior was further clarified by examining the fluctuating energy components: oblate and prolate particles differ in how this energy is distributed and expressed in their dynamics. 
Moreover, slightly aspherical particles at low friction exhibit non-monotonic behavior, which we linked to enhanced rotational kinetic energy fluctuations. \\
In addition, flat particles were found to exhibit more pronounced fabric anisotropy compared to their elongated counterparts. 
Finally, our analysis of power dissipation showed that it tends to cluster along the major axis, aligned with the flow, where slip events are more frequent, consistently across all particle shapes.

This work can be extended in several directions. 
First, the relationship between fabric anisotropy and stress transmission in non-spherical particle assemblies could be further examined within the framework proposed by Srivastava et al. \cite{srivastava2021viscometric}. 
Additionally, incorporating frictional effects into the anisotropic rheological model proposed by Nadler et al. \cite{nadler2021anisotropic} or the correlation lengths into extended kinetic theories \cite{berziDenseShearingFlows2022} could enhance predictive capabilities for both natural and industrial applications.\\
Future work should explore the connection between the dissipation-based regimes identified in our study and the stress-based regimes (e.g., rate-dependent vs. rate-independent) discussed in the literature \cite{campbell2002granular, campbellStresscontrolledElasticGranular2005, berziGranularFlowsKinetic2024}.

Another key avenue for investigation is the role of vortices and clusters in shear-driven particulate flows, as discussed in \cite{oyamaAvalancheInterpretationPowerLaw2019, rognon2021shear, zou2022microscopic, athani2024unifying}.
The formation and dynamics of vortical structures could significantly impact shear banding and localized dissipation, necessitating a deeper physical understanding. 
Moreover, extending our analysis to dense suspensions \cite{trulsson2017effect} will help bridge the gap between granular and suspension rheology, elucidating the transition between friction-dominated and hydrodynamic-dominated flow regimes. 
Lubrication forces may become dominant at lower viscous numbers, especially for flat particles which enhance lubrication effects.
Finally, a systematic study should be conducted on non-convex, potentially interlocking particles \cite{saint2011rheology}.
A deeper understanding of these interactions could inform the design of polycatenated materials \cite{zhou20253d} with novel particle architectures, ultimately improving control over their flow properties.

\section*{Acknowledgements}
We thank Professor Ali Ozel for sharing the preliminary input scripts on simple shear under quasi-static conditions, which contributed significantly to the work presented in \cite{angus2020calibrating}.
We also acknowledge Lucas Fourel for expert assistance with visualization software and insightful advice on data presentation.
This project has received funding from the European Union’s Horizon 2020 research and innovation programme under the Marie Skłodowska-Curie grant agreement No 945363.
J.B. and J.-F.M. further appreciate the financial support from Bühler AG and the valuable discussions with Clément Zemerli and Richard Vonlanthen.
M.T. acknowledges funding from the Swedish Research Council under grant No. 2021-04997.

\section*{Data Availability}
Data to support these findings is available on Zenodo \cite{bilotto_2025_15798937}.
Codes to run the simulations and the analysis are available on GitHub at \url{https://github.com/JBil8/Simple_shear_granular_spheroids.git}.

\appendix

\section{Contact interaction}
\label{app:contact}

As stated in Section \ref{sec:simulations}, all simulations use the Hertz-Mindlin contact mechanics model, with damping in the normal direction.
For each contact the normal and tangential force are computed with the following formulas:
\begin{align}
	\mathbf{F}_n &= (k_n \delta_n - \gamma_n \dot{\delta}_n) \mathbf{n}^c \\
        \mathbf{F}_t &= \min((k_t \boldsymbol{\delta}_t, \mu_p F_n \mathbf{u}_t/|\mathbf{u}_t|) \,    
\end{align} 
where $\delta_n, \boldsymbol{\delta}_t$ are the normal and tangential overlap, $\dot{\delta}_n, \, \mathbf{u}_t$ are the relative velocity of the particles at the contact point in the normal and tangential direction respectively, $k_n, \, k_t$ are the normal and tangential stiffnesses, and $\gamma_n$ is the normal damping coefficient.
The tangential overlap is computed as
\begin{align}
    \bm{\delta}'_t = \int_{T_c} \bm{u}_t dt \approx \sum_{i=0}^{n_{stc}} \bm{u}_t^i \Delta t \, ,
\end{align}
where $T_c$ denotes the total contact duration, and 
$n_{stc}$ is the number of discrete timesteps during which contact occurs.
Additionally, at each time step the computed shear is projected onto the tangential plane of the current contact.

When slip occurs, \emph{i.e.} $|\mathbf{F}_t| = \mu_p |\mathbf{F}_n|$, the tangential overlap is rescaled to $\bm{\delta}_t = \mathbf{-F_t}/\mu_p$.
The contact stiffness and damping coefficient are computed as follows:
\begin{align}
    k_n &= 4 E^* \sqrt{R^* \delta_n}\\
    k_t &= 8 G^* \sqrt{R^* \delta_n} \\
    \gamma_n &= -2 \sqrt{5 \over 6} {ln(e) \over \sqrt{ln^2(e)+\pi^2}} \sqrt{2E^* m^*} (R^*\delta_n)^{1/4}
\end{align}
with $E^*, \, G^*, \, R^*, \, m^*$ the equivalent Young modulus, shear modulus, contact radius and mass for the contact between two particle with identical material properties, defined as:
\begin{align}
    E^* & = {E \over 2 (1-\nu^2)} \\
    G^* & = { E \over 4 (2-\nu)(1+\nu)} \\
    {1 \over R^*} & = {1 \over R_1^{eq}} + {1 \over R_2^{eq}} \\
    {1 \over m^*} & = {1 \over m_1} + {1 \over m_2} \, .
\end{align}
$R^{eq}$ is the equivalent contact radius for a particle which is computed as the inverse of the gaussian curvature coefficient, $\kappa_g$, 
\begin{align}
    R^{eq} = {1 \over \kappa_g} = {1 \over \sqrt{K}} \,
\end{align}
where $K$ is the gaussian curvature at the contact point.
For a spheroid, in the local coordinate system the gaussian curvature is simply given by:
\begin{align}
    K = {c^6 \over
    [c^4 +(a^2-c^2)\mathcal{Z}^2]^2} \, .
\end{align}

The Hertzian contact time and the Rayleigh time of Section \ref{sec:simulations} are defined as:
\begin{align}
    T_H = &  2.94 \left( 5 \sqrt{2} \pi \rho_p \over 4 E^*
    \right)^{2/5} {a_{\text{min}} \over \dot{\gamma} d_{eq} }  \, 
    \label{eq:t_hertz}\\
    T_R = & {\pi a_{\text{min}} \over 0.8766+0.163 \nu } \sqrt{\rho_p \over G} \, 
    \label{eq:t_rayleigh}
\end{align}
where $G$ is the shear modulus and we assumed the typical impact velocity to be of the order of the one imposed by the shear flow ($\dot{\gamma} d_{eq}$), and $a_{\text{min}}$ is the smallest dimension of the smaller grains.

Each spheroid is described in global coordinates with its corresponding level set function $G(\mathbf{x})$.
The overlap between two particles is computed by solving the following optimization problem:
\begin{align}
    &\min(G_1(\mathbf{x}) + G_2(\mathbf{x}))\\
    &\text{subject to } G_1(\mathbf{x})=G_2(\mathbf{x}) \, .
\end{align}
With a few Newton-Rapson iterations, both the contact point and the total overlap are found, which then allows the computation of the interaction force as described above. 
More details can be found in \cite{podlozhnyukEfficientImplementationSuperquadric2016}.

\section{Dissipation computation}
\label{app:dissipation}

\begin{figure}
    \centering
    \includegraphics[width=\linewidth]{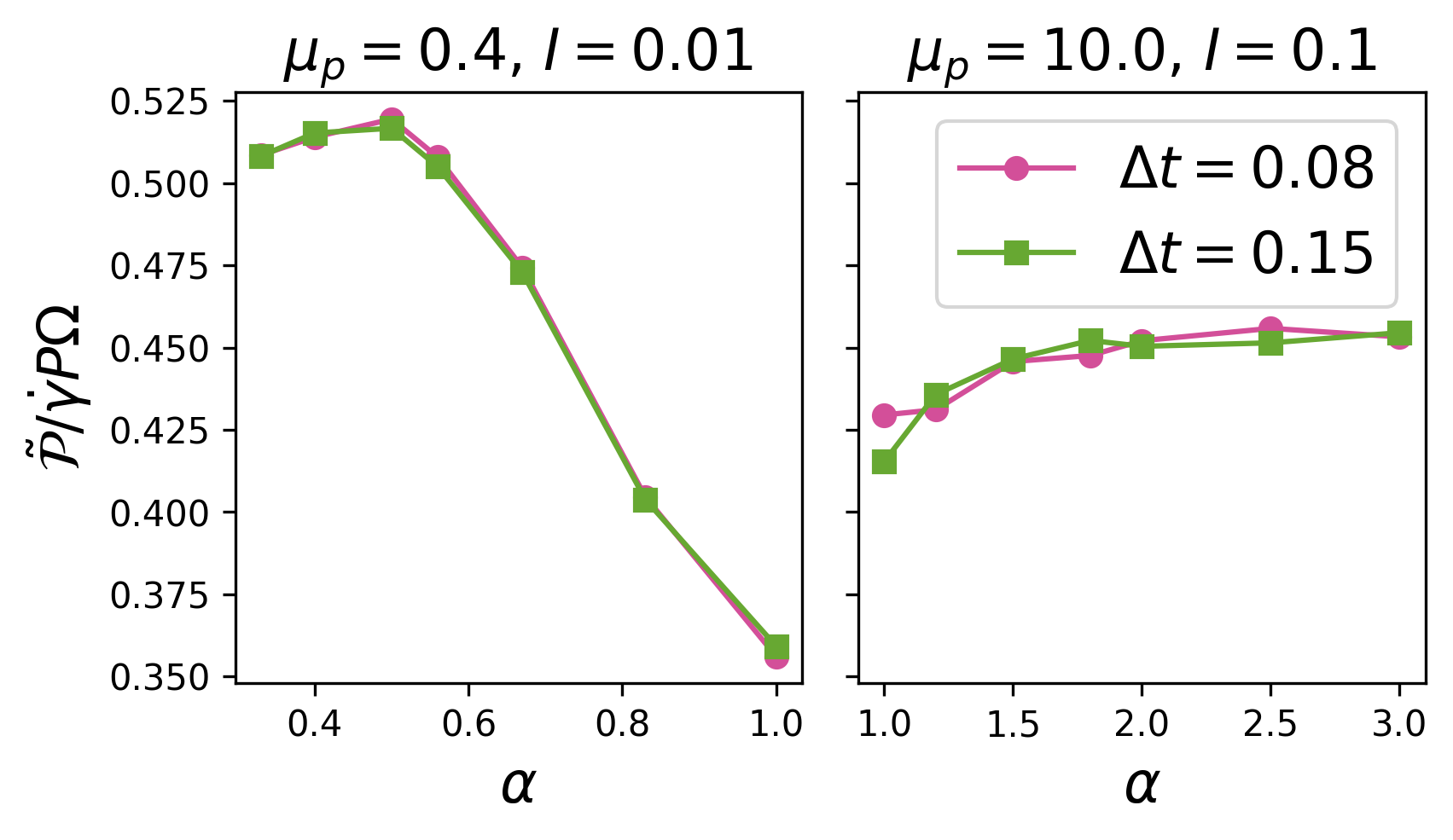}
    \caption{Influence of time step on the computed dissipated power using the sum of Equations \ref{eq:dissipation_normal} and \ref{eq:dissipation_tangential}.
    (Left) data for  $I=0.01$ and $\mu_p=0.4$, corresponding to the sliding regime.
    (Right) data for
    $I=0.1$, $\mu_p=10$, corresponding to the rolling regime.
    The timestep is expressed as a fraction of the smallest between $T_H$ and $T_R$, see Equations \ref{eq:t_hertz} and \ref{eq:t_rayleigh} respectively.}
\label{fig:dissipation_time_step}
\end{figure}
We used a velocity Verlet integration scheme to perform the numerical integration of the equations of motion in time.
To accurately compute the dissipation during a time step, one should use knowledge about the state of the system at the start and middle of the time step \cite{mayrhofer2017energy}.

However, we only dump the information at the end of selected time steps and therefore approximate the average power dissipation by taking an ensemble average.
The normal dissipation during a time step is given by \cite{mayrhofer2017energy}:
\begin{align}
    \mathcal{P}_n 
    &= \sum_{N_c}\left[
    \frac{\mathbf{F}^{\gamma}_{n,i}}{2} \cdot 
    \left( \mathbf{u}_{n,i} + \frac{1}{4m*} \mathbf{F}_{n,i} \right)
    + \right. \nonumber \\
    &\quad \left.
    \frac{\mathbf{F}^{\gamma}_{n,i+1}}{2} \cdot 
    \left( \mathbf{u}_{n,i+1/2} + \frac{1}{4m*} \mathbf{F}_{n,i+1} \right)
    \right]      \nonumber \\
    &\approx \sum_{N_c} \mathbf{F}_{t,i} \cdot \mathbf{u}_{t,i} \, ,
    \label{eq:dissipation_normal}
\end{align}
and the last equality is satisfied for small enough time steps, so that averaging at half and full time step gives a similar result to the full time step only.

Frictional dissipation occurs only at sliding contacts and during a time step of size $\Delta t$ it amounts to:
\begin{align}
    \mathcal{P}_n 
    = \sum_{N_s} \mathbf{F}_{t,i} \cdot
    \left(
        \frac{\bm{\delta}_{t,i} + \mathbf{F}_{t,i}/k_t}{\Delta t}
    \right)
    \approx 
    \sum_{N_s} \mathbf{F}_{t,i} \cdot \mathbf{u}_{t,i}
    \label{eq:dissipation_tangential}
\end{align}
where the second equality is satisfied for small enough $\Delta t$ so that slip only occurs at the beginning of the time step and the tangential velocity can be considered constant during the interval.
We verified both Equations \ref{eq:dissipation_normal} and \ref{eq:dissipation_tangential} this by halving the time step size.
In particular Equation \ref{eq:dissipation_normal} is tested for those simulations with the highest fluctuations $\mu_p =10$, $I=0.1$, corresponding to dominant normal dissipation, while Equation \ref{eq:dissipation_tangential} is tested for $\mu_p =0.4$, $I=0.01$, where frictional dissipation is dominant.
Figure \ref{fig:dissipation_time_step} shows results within statistical error, implying that the chosen time step is sufficiently small.

\section{Finite Size effects}
\label{app:finite_size}

\begin{figure}
    \centering
    \includegraphics[width=\linewidth]{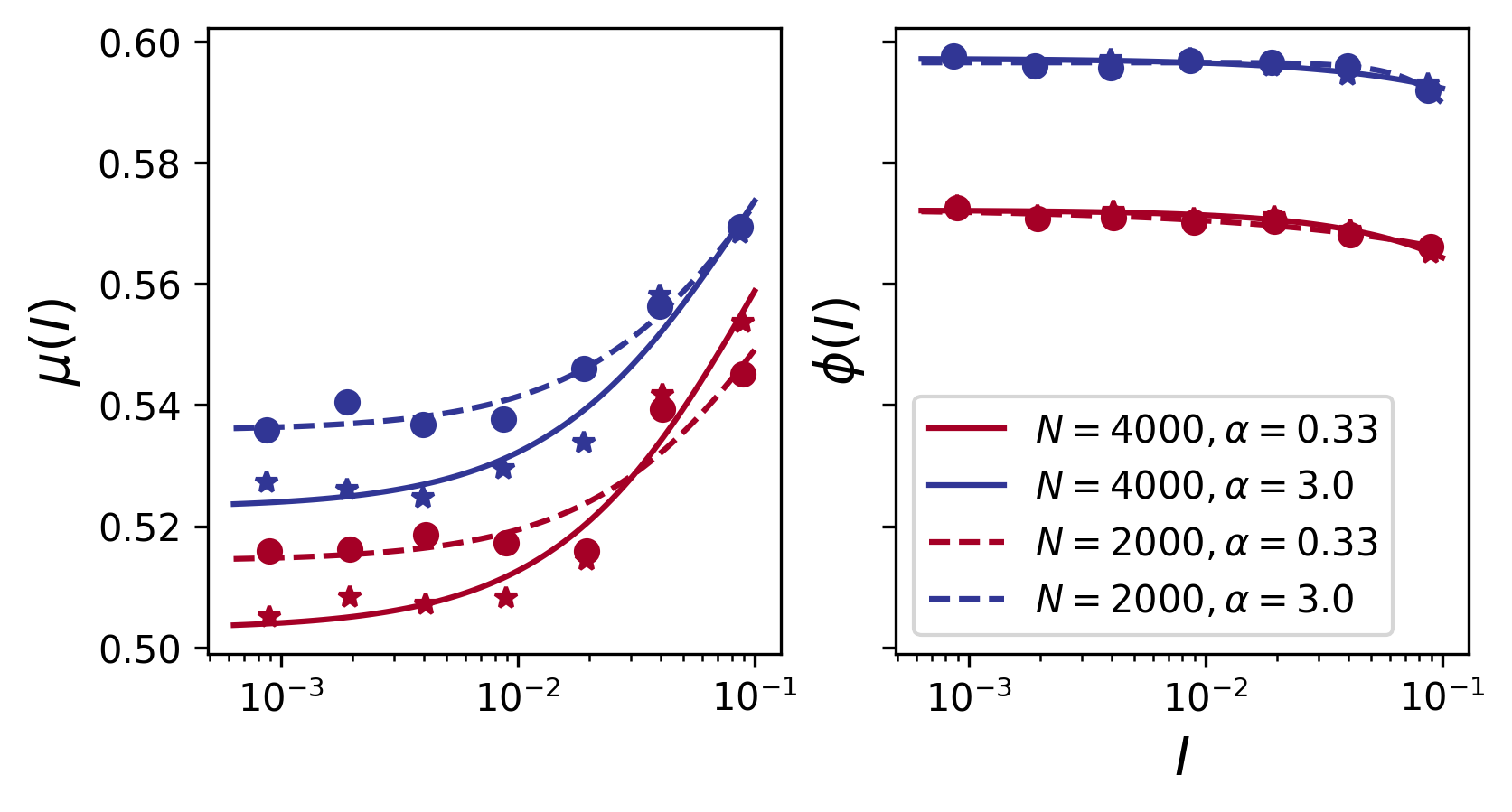}
    \caption{Finite size effects for the rheological curves for the two extreme aspect ratios, and $\mu_p=0.4$, (left)  $\mu(I)$ and (right) $\phi(I)$ }
    \label{fig:finite_size_effects}
\end{figure}
We perform some additional simulations with $N=4000$ particles in the simulation box, for which we report the rheological curves for $\mu(I)$ and $\phi(I)$ in Figure \ref{fig:finite_size_effects}.
We find that the volume fraction remains unchanged, while the measured friction coefficient is slightly lower at low inertial numbers, but the difference is smaller than $2\%$.
Studies of bigger systems would anyway be valuable to compute correlation lengths, especially at low inertial numbers when the correlation length increases.

\section{Normal stress difference}
\label{app:normal_stresses}

\begin{figure}
    \centering
    \includegraphics[width=\linewidth]{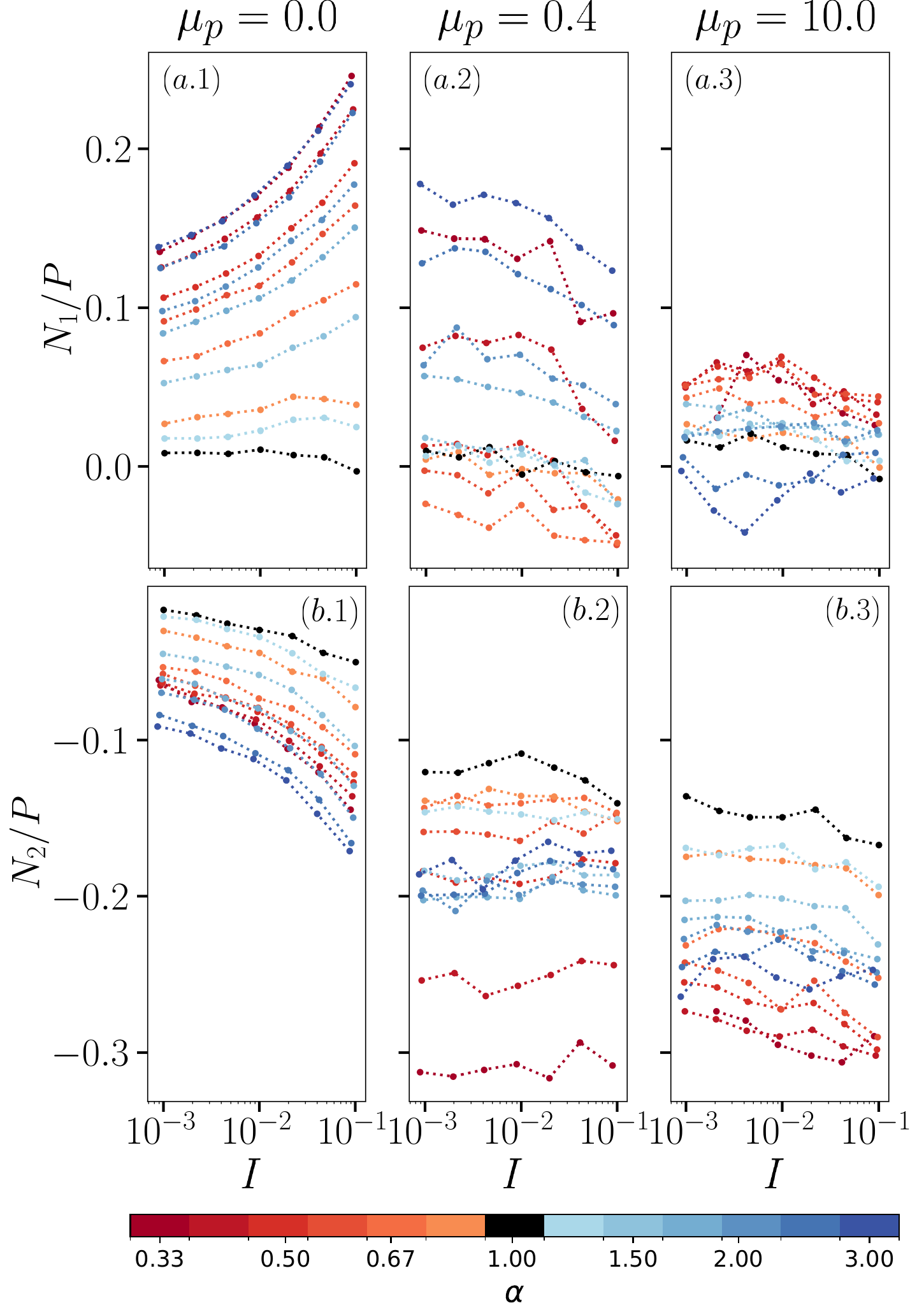}
    \caption{First (top) and second (bottom) normalized normal stress differences as a function of inertial number.
    }
    \label{fig:fig:normal_stresses_diffusion}
\end{figure}

Another characteristic of granular materials is their display of normal stress differences \cite{trulsson2018rheology, nagy2017rheology, nagy2020flow}.
We define the first normal stress difference, $N_1 = \sigma_{xx}-\sigma_{yy}$, the second normal stress difference, $N_2 = \sigma_{yy}-\sigma_{zz}$.\\
Looking at Figure \ref{fig:fig:normal_stresses_diffusion} the first normal stress difference shows an increasing trend with $I$ for $\mu_p=0$ and a slightly decreasing one in the frictional case.
We report $N_1<0$ at $I>10^{-2}$ for slightly oblate particles $0.6 < \alpha < 0.83$ at $\mu_p=0.4$, even though no shear band is observed.
\\
The second normal stress difference is always negative and shows a declining trend with respect to $I$.
While more elongated prolate particles experience more compression along the vortex lines \cite{maklad2021review} in the frictionless case, flatter particles do so in the frictional one.
Intuitively, flattened particles experience higher compressive forces on their major axis (due to the smaller local curvature).
In the frictional case, the Coulomb's frictional cone is such that the resulting force between spheroids in two adjacent layers has a small $z$ component, whereas in the frictionless case the resulting force is simply oriented along the normal. 
Interestingly, in the highly frictional limit $N_2$ is greater than in the $\mu_p=0.4$ case for $\alpha<0.5$, in contrast to the monotonic trend for the other values. 

\section{Particle circulation and diffusive behaviour}
\label{app:diffusion}

\begin{figure}
    \centering
    \includegraphics[width=\linewidth]{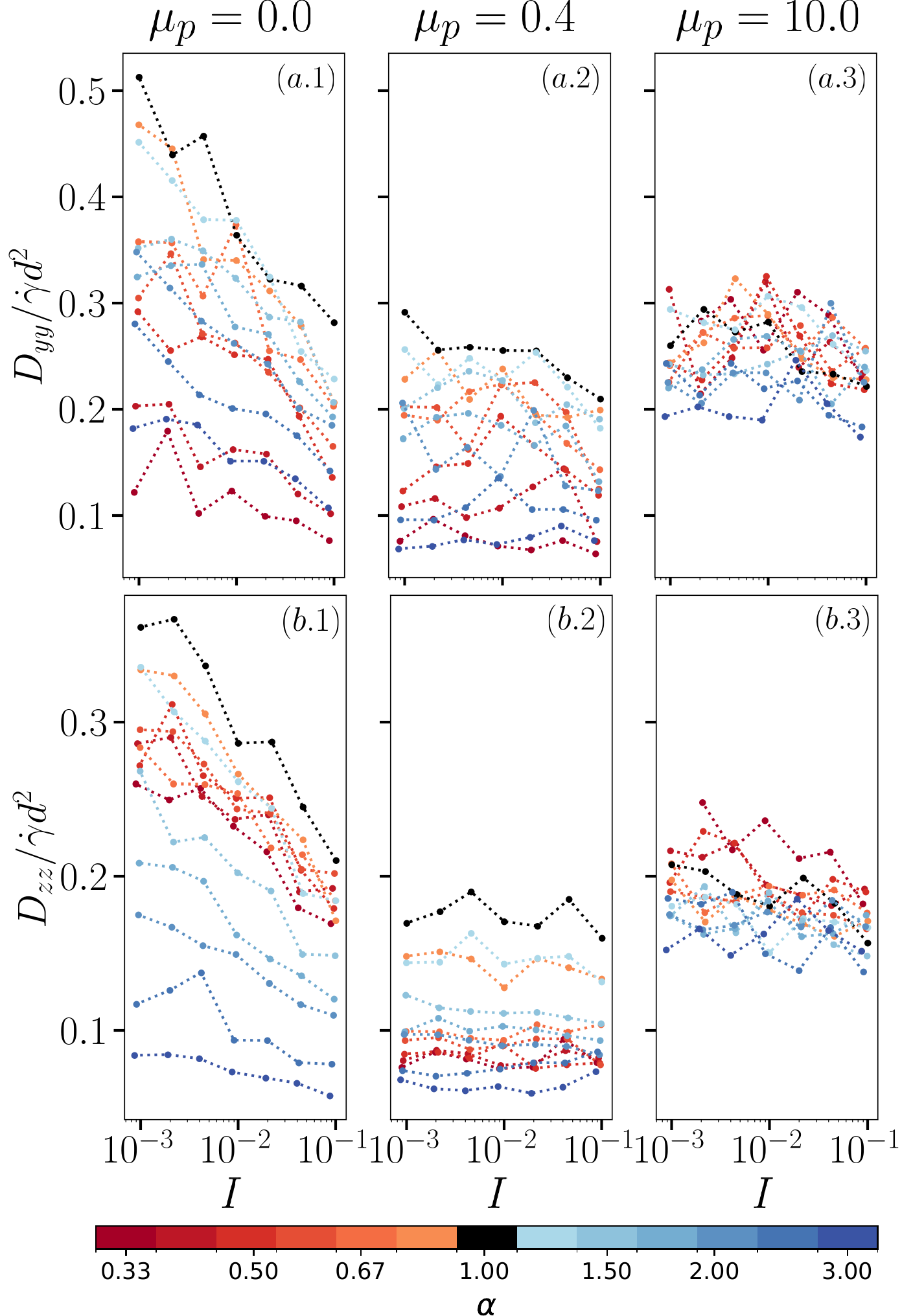}
    \caption{Gradient (top) and vorticity (bottom) diffusion coefficient, normalized by $\dot{\gamma} d_{eq}^2$, as a function of inertial number.  }
    \label{fig:diffusion}
\end{figure}

The diffusion coefficient is extracted from the mean squared displacement using the Einstein relation, $D_{k} = \mathrm{lim}_{t \to \infty} \langle k^2 \rangle / 2t  $, with $k = y, z$, and numerically estimated via least-square fitting.
As shown in Figure \ref{fig:diffusion}, for frictionless particles, the diffusion coefficient exhibits a slightly decreasing trend with the inertial number, while for frictional cases, it remains nearly constant. 
Unlike previous studies \cite{kharel2017vortices, rognon2021shear, athani2024unifying}, our data do not follow the expected scaling $D_{yy}\sim \dot{\gamma} d^2 I^{-1/2}$, likely due to finite size effects, where approximately 15 particles span the domain in the $y-$direction, constraining vortex formation to the system size.
Even in the case of frictionless spheres, the best-fit power law decay is around $D_{yy}\sim \dot{\gamma} d^2 I^{-1/8}$.
It is likely that the observed reduction is a distinctive feature of 3D systems—analogous to vortex stretching in Newtonian fluids, which occurs exclusively in three dimensions \cite{oyamaAvalancheInterpretationPowerLaw2019}.

Despite some noise in the data, elongated and flat particles generally exhibit lower diffusion compared to spheres in all cases except for the highly frictional rolling regime.
In most cases, the shear-induced alignment of elongated particles creates a cage-like structure that restricts their mobility, thereby reducing diffusion \cite{cai2021diffusion}. 
Interestingly, oblate particles exhibit greater diffusion in the vorticity direction ($z$) compared to prolate ones, suggesting a weaker caging effect along this axis.

For frictional particles, we also observe entanglement and rotational motion, similar to what was reported for flat disks and elongated rods under Lees-Edwards boundary conditions \cite{guo2013granular}.
In our simulations, where simple shear is achieved by physically tilting the simulation box, this phenomenon primarily occurs when the box tilt is near zero (see movie in supplemental material \cite{bilottosupp}).
The circulation arises due to the absence of box tilt, allowing particles to move coherently along $y$ since they are reinserted at approximately the same $x$-coordinate when crossing the periodic boundary.
This effect is further supported by the mean square displacement in $y$, which exhibits periodic peaks over time (data not shown).
Additional simulations with $L_x >7L_y$ confirm a similar behavior.
The observed coherent motion is linked to the vortex structures described in \cite{rognon2021shear}, which in 3D can span the whole domain size at high densities, unlike the correlation lengths \cite{oyamaAvalancheInterpretationPowerLaw2019}.

\section{Spatial distribution of contacts}
\label{app:global_contact_distribution}

We bin forces in the global reference frame $(x,y,z)$ by the angle they make with $x$ axis in the $xy$ plane.
We compute the probability of having a force at a certain angular location, $\rho$, which satisfies the following normalization condition:
\begin{align}
    \int_0^{2 \pi} \rho(\beta) d\beta = 1 \, ,
\end{align}
and the average force, conditional to a force being present $\overline{F}$, projected in the $xy$ plane.
\begin{figure}
    \centering
    \includegraphics[width=\linewidth]{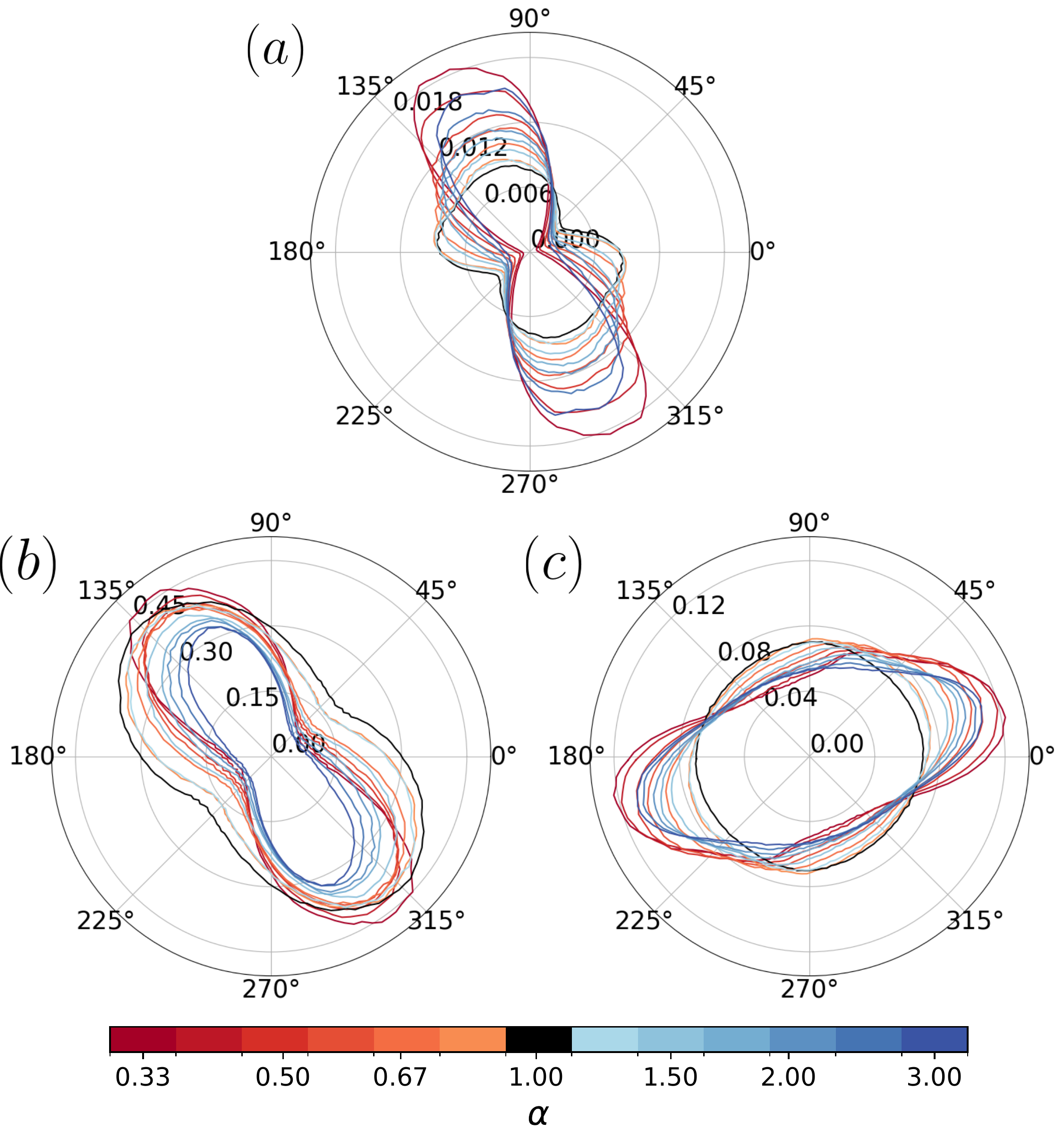}
    \caption{Angular distribution of contacts normal (a) direction in the global reference frame, i.e. angle$=0^\circ$ for a contact along $x$, $90^\circ$ for a contact along $y$.
    Normalized conditional force $\overline{F}_n / P A_p$ (b) and $\overline{F}_t / P A_p$ (c).}
    \label{fig:global_contact}
\end{figure}

In Figure \ref{fig:global_contact}, the density distribution shows two peaks at $\beta = 0$ and $\beta \approx - \pi /3$ for spherical particles, similarly to circles in 2D \cite{da2005rheophysics}.
Particle's eccentricity makes the peak in the flow direction shift towards the compression direction, increasing the anisotropy of the stress.
This means that both flat and elongated particles tend to have very few contacts in the flow direction.
Remarkably, oblate grains display more peaked distributions at high $|r_g|$ compared to prolate ones.\\
Similarly, $\overline{F_n}$ is higher in the compression direction for spheres, but shift clockwise due to elongation (see Figure \ref{fig:global_contact}).
Therefore, despite most particles being aligned at a specific angle and most contacts per unit area occurring on the major axis, the most likely direction of contact remains along the compression direction.\\
The tangential forces $\overline{F}_t$ act predominantly in the gradient direction for spheres, but shift at an angle toward the flow direction for flat and elongated particles.

\section{Decomposition of shear stress in normal and tangential component}
\label{app:stress_decomposition}
\begin{figure}
    \centering
    \includegraphics[width=0.75\linewidth]{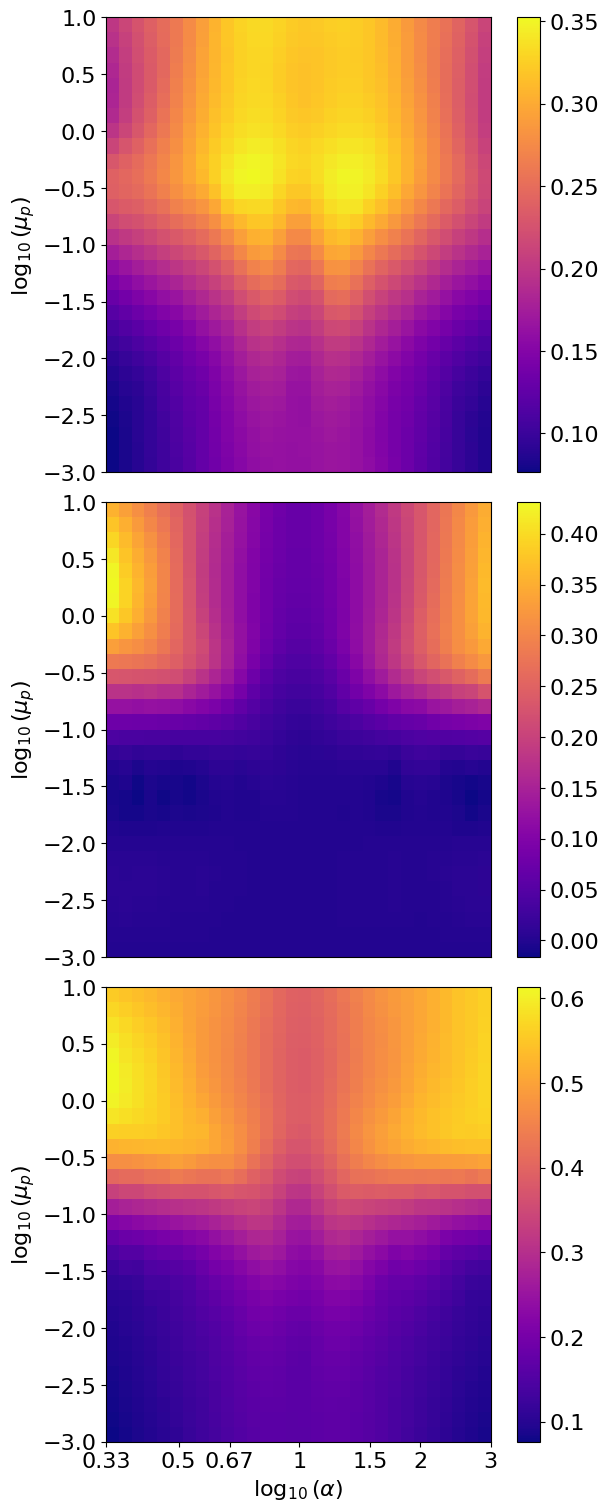}
    \caption{Heatmap of normalized shear stress $\sigma_{xy}/P$ as a function of $\mu_p$ and $\alpha$, for $I\approx0.01$.
    (Top) normal component of the shear stress, (Middle) tangential component, (Bottom) total shear stress.
    }
    \label{fig:heat_map_shear_stress}
\end{figure}
Following similar reasoning to \cite{zou2023particle}, we split the virial stress into the sum of the one caused by normal forces and the one due to tangential ones:
\begin{align}
    \boldsymbol{\sigma} =& \boldsymbol{\sigma}_n + \boldsymbol{\sigma}_t\\
    \sigma_{ij}^n =& {1 \over 2\Omega} \sum_{k, l \in V}^N  (r_{j}^{(k)}- r_{j}^{(l)})F_{i}^{n_{lk}} \\
    \sigma_{ij}^t =& {1 \over 2\Omega} \sum_{k, l \in V}^N  (r_{j}^{(k)}- r_{j}^{(l)})F_{i}^{t_{lk}} \, .
\end{align}

Figure \ref{fig:heat_map_shear_stress} displays data for different shapes and microscopic friction.
The normal component of the shear stress increases up to moderate values of friction, but decreases for higher values.
Interestingly, the maximum $\sigma_{xy}^n$ is found for particles shapes that deviate slightly from spheres, at intermediate values of $\mu_p$, which might be another way to interpret the non-linearity of the dissipation regime diagram at small eccentricity.\\
The tangential component of the shear stress generally increases as $\mu_p$ is increased, and becomes dominant at $|r_g|>0.33$ higher values for both flat and elongated particles.
The total shear stress and its tangential component are maximum for flat particles at $\mu_p \approx 1.0$, and not in the high friction limit.
We therefore conclude that, given a certain shape, the maximum shear strength is found at finite values of $\mu_p$, those for which the optimal alignment of particles is achieved.

\bibliography{references, references_zotero}

\end{document}